\documentclass[
aps, 
reprint,superscriptaddress,floatfix,
longbibliography
]{revtex4-2}

\usepackage{amsthm}
\usepackage{amsmath,amssymb} 
\usepackage[utf8]{inputenc}
\usepackage{graphicx}
\usepackage{dcolumn}
\usepackage{bm}
\usepackage{dsfont}
\usepackage[german, english]{babel}
\usepackage{xcolor}
\usepackage{mathtools}
\usepackage{comment}
\usepackage{tabularx}
\usepackage{array}

\usepackage{mathtools} 
\usepackage{xspace}

 \usepackage{hyperref}
 \hypersetup{
colorlinks=true,
urlcolor= blue,
citecolor=blue,
linkcolor= blue,
breaklinks=true
}

\newcommand{\tauopt}{\tau_{\text{opt}}}
\newcommand{\taup}{\tau_{+}}
\newcommand{\alphaopt}{\alpha_{\text{opt}}}

\renewcommand{\arraystretch}{1.5}

\addto\captionsenglish{}
\addto\captionsenglish{}

\newcolumntype{Y}{>{\centering\arraybackslash}X}

\begin{document}

\title{
When it pays to teach: a population threshold for dedicated teaching 
}

\author{Hirotaka Goto}
\affiliation{Graduate School of Advanced Mathematical Sciences, 
Meiji University, 4-21-1 Nakano, Tokyo 164-8525, Japan}

\author{Joshua B. Plotkin}
\affiliation{
Center for Mathematical Biology, University of Pennsylvania, Philadelphia, PA 19104
}
\affiliation{
Department of Biology, University of Pennsylvania, Philadelphia, PA 19104
}

\date{June 30, 2025}

\begin{abstract}
Teachers hold a prominent place in modern societies, particularly where education is compulsory and widely institutionalized. This ubiquity obscures an underlying question: why do societies designate certain individuals exclusively for the instruction of others? This question is especially enigmatic for dedicated teachers, who invest their labor in cultivating others' skills but do not directly participate in the productive activities for which their students are being trained. To address this puzzle, we develop a simple, mathematically tractable model of teaching and learning in a population with a shared goal. We identify a tradeoff between the size of the workforce and its collective level of expertise; and we analyze the optimal proportion of a population that should serve as teachers across a wide range of scenarios. We show that a population must exceed a critical size before it is beneficial to allocate anyone as a dedicated teacher at all. Subsequently, the peak demand for teachers is achieved at an intermediate population size, and it never surpasses one half of the population. For more complicated tasks, our analysis predicts the optimal allocation of teachers across different levels of expertise. The structure of this teacher allocation is more complex when the population size is large, in agreement with the general size--complexity hypothesis. 
Our account lays a foundation for understanding the adaptive advantage of dedicated teachers in both human and non-human societies.
\end{abstract}

\maketitle

\section{\label{sec:introduction} Introduction}

\begin{quote}
    Those who can, do; those who can't, teach. 
\end{quote}
George Bernard Shaw's famous quip casts teaching as a superfluous activity of the inept. Nonetheless, since ancient times complex societies devote some individuals to teaching, presumably in return for increased expertise among a larger class of students and eventual practitioners. Societies therefore seem to value teaching for the enhanced productivity it can provide to the population at large.

Teaching is key to the transmission of ``opaque culture''---that is, complex traditions or skills that cannot easily be inferred through direct observation alone \cite{Garfield2025}. 
Most contemporary societies have instituted systems of education, devoting a portion of their population purely to the instruction of others. 
We refer to this as \emph{dedicated teaching} to indicate that such teachers do not themselves engage in any directly productive activity.
The long-standing practice of dedicated teaching, whether institutionalized or not, suggests it may offer an advantage to individuals, kin, or groups \cite{Allchin2024}. 

In fact, dedicated teaching is not unique to humans; it is widely observed in nature as well. At the molecular scale, chaperones assist other proteins to fold or unfold properly, but they do not themselves take part in functional enzymatic activity within the cell \cite{Macario2007, Moller2024}. 
In mammals, post-reproductive individuals can impart valuable information across generations. The \emph{grandmother hypothesis} \cite{Cant2008, Coxworth2015, Kim2012increased, Kachel2011} posits that the post-reproductive female lifespan has evolved for alloparental care, contributing to the fitness of offspring \cite{Hawkes2004, Blell2018, Hawkes1998, Lahdenpera2018, Hagen2025}. Several lines of empirical evidence show that interaction with grandmothers improves survival of grand-offspring in both human and non-human species, including killer whales \cite{Nattrass2019}, elephants \cite{Lee2016, Lahdenpera2016}, and giraffes \cite{Muller2022}.  
Social and biological underpinnings aside, there is widespread empirical evidence of dedicated teaching in natural populations. 

Given the prevalence of teaching in nature, Bernard Shaw's quip presents a puzzle. Dedicated teachers, by definition, provide no productivity or direct benefit to a group, which raises several key questions: when does dedicated teaching provide a net benefit to the group even at the cost of reducing the workforce? What proportion of a population should dedicate itself to teaching others? And, critically, how does a group's size modify the optimal proportion of dedicated teachers? Related questions include how teaching efforts should be allocated when students differ in their skill levels---beginners or high achievers? 

Here we pose and study these questions by developing a tractable mathematical model that can account for the emergence of a dedicated teacher class within a population. We develop a model in which individuals can either produce or teach, and the population shares the rewards of production. 
Our model assumes that dedicated teachers help learners to improve their skill at an activity that is beneficial to the group. We refer to this productive activity as ``hunting," although we use this term only as an illustrative example of any broadly useful skill. In our model, a hunter can alternatively increase their skill through self-learning rather than by tutelage; and an expert may forget their skill at some rate. Teachers, by contrast, never hunt and therefore do not contribute directly to group productivity. 

The model encapsulates a basic trade-off: the more individuals are allocated to be teachers, the more efficiently others will acquire expertise, but the fewer productive individuals (hunters) are available.
We analyze this trade-off by identifying the proportion of teachers that maximizes overall group productivity. 
We find that a population must reach a minimum size before it is beneficial to allocate anyone to the teacher role whatsoever.  Thereafter, the optimal proportion of teachers is maximized for a population of intermediate size, and the maximum can never exceed \( 50 \)\% of the population. We also obtain a simple estimate for the peak educational demand when expertise provides only a small advantage. Finally, we extend our model to include three levels of expertise instead of two, i.e., when the task becomes more complex.
Our results suggest that optimal teaching usually has two turning points---determining when to dedicate anyone \emph{at all} to teaching and when to start educating \emph{non-beginners}---which unfold sequentially as the population size grows. 
We conclude by discussing our work as an elementary framework for understanding the adaptive advantage of a ``teacher class'' in human and non-human societies.

\section{Model}

\subsection{A Model of Teaching and Learning}

\begin{figure}[t]
    \centering
    \includegraphics[width=0.5\linewidth]{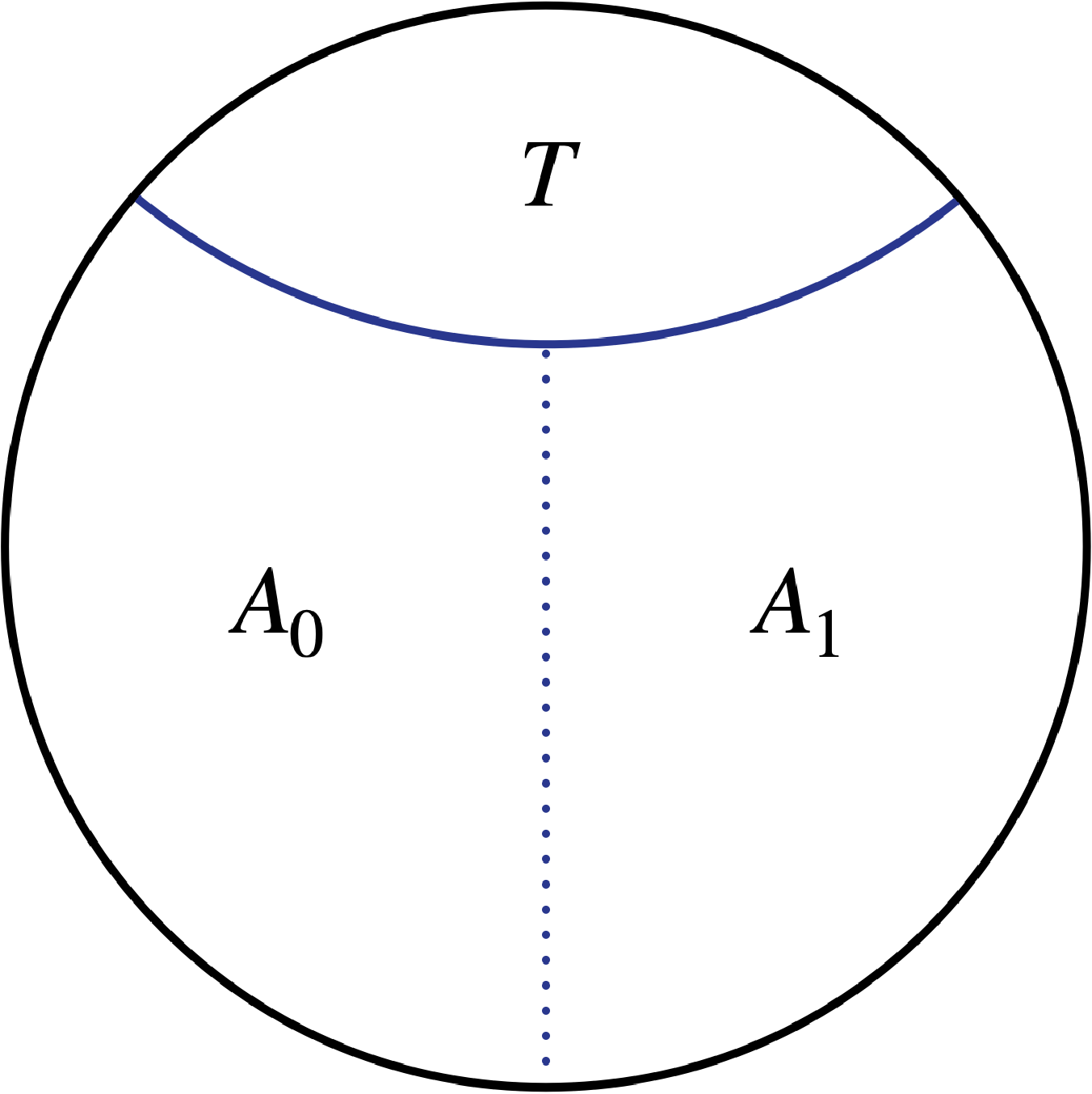}
    \caption{Schematic illustration of the population structure with amateur hunters \( A_0 \), expert hunters \( A_1 \), and dedicated teachers \( T \). Both the total population size, \( n \), and the fraction of teachers, \( \tau \), are fixed; and these in turn determine the equilibrium number of amateur and expert hunters.
    } 
    \label{fig:schematic_illustration}
\end{figure}

Our basic model posits three mutually exclusive types of individuals: 
amateur hunters, expert hunters, and teachers, which we denote by \( A_0 \), \( A_1 \), and \( T \), respectively 
(see Fig.~\ref{fig:schematic_illustration} for a schematic illustration). 
Amateur hunters engage in hunting at a rate of \( h_0 \), whereas expert hunters hunt at a higher rate \( h_1 >  h_0\). 
Teachers, on the other hand, do not hunt at all but train amateur hunters to become expert at hunting, which we call \emph{education}. 
Amateur hunters can also learn on their own to become expert hunters, which we refer to as \emph{self-learning}. 
Finally,
expert hunters---either self-taught or educated---may forget their expertise at some rate, 
reverting to the amateur status. 
For simplicity, all interactions between individuals are mean-field and occur in a well-mixed population of a given size \( n \). 
We focus our analysis on the following question: 
how many teachers should be present in a population of size \( n \) to maximize the collective hunting productivity?

\begin{table}[t!]
    \centering
    \begin{tabular}{cl}
        \hline \hline
        Symbol & Description \\
        \hline 
        \( A_0 \) & Amateur hunter \\
        \( A_1 \) & Expert hunter \\
        \( T \)  &  Teacher \\  
	\( n \)  &  Total population size \\
	\( n_0 \)  &  Number of amateur hunters \\
	\( n_1 \)  &  Number of expert hunters \\  
	\( n_T \)  &  Number of teachers \\  
	\( \lambda \) & Self-learning rate \\ 
	\( \mu \) & Education rate \\ 
	\( f \) & Forgetting rate \\
	\( h_0 \) & Amateur hunting rate \\
	\( h_1 \) & Expert hunting rate \\ 
	\( \tau \) & Teacher-to-population ratio \\ 
        \( F \) & Per-capita productivity \\
        \( \eta \) & Self-learning efficiency (\( \eta = \lambda / f \))\\
        \( \nu \) & Education efficiency (\( \nu = \mu / f \)) \\ 
        \( r \) & Advantage of expertise (\( r = h_1 / h_0 \)) \\
        \hline \hline
    \end{tabular} 
    \caption{Notation for the baseline hunter-teacher model.}
    \label{tab:notations}
\end{table}

To address this question in a well-defined way, 
we model the three independent processes outlined above using the following dynamical system: 
\begin{align} \label{eq:baseline_model}
\begin{cases}
\dot n_0 = f n_1 - \left( \lambda + \mu n_T \right) n_0, \\
\dot n_1 = \left( \lambda + \mu n_T \right) n_0  - f n_1, 
\end{cases}
\end{align}
where \( n_0(t) \) and \( n_1(t) \) respectively denote the numbers of amateur and expert hunters at time \( t \), 
and \( n_T \) is the number of teachers in the population, which is set exogenously and assumed constant. 
The parameters 
\( f \), \( \lambda \), and \( \mu \) represent the rates of \emph{forgetting}, \emph{self-learning}, and \emph{education}, respectively. 
The resulting per-capita hunting productivity \( F \) is given by 
\begin{align} 
    \label{eq:baseline_productivity}
    F = \frac{n_0 h_0 + n_1 h_1}{n}, \quad 0 < h_0 < h_1, 
\end{align}
where we substitute the equilibrium frequencies of amateur and expert hunters from Eq.~\ref{eq:baseline_model} into \( n_0 \) and \( n_1 \). 
Finally, let \( \tau \) denote the proportion of teachers (\( 0 \leq \tau \leq 1 \)), so that 
\( n_T = \tau n \) and 
\( n_0(t) + n_1(t) = (1-\tau)n \) for all time \( t \). 
We refer the reader to Table~\ref{tab:notations} for a complete list of variables and parameters, which are all positive by definition. 

Now that we have formulated the model, we start with some simple qualitative observations before delving into an analysis of the optimal proportion of teachers.
Note that a population consisting entirely of teachers is never optimal because nobody hunts (i.e., \( F = 0 \)). 
For a given total population size, a large number of dedicated teachers improves 
hunting efficiency at the expense of the number of individuals who actually hunt; 
by contrast, a population with a small number of dedicated teachers suffers from inefficient hunting in exchange for large workforce. Taken together, there is 
a trade-off in allocating dedicated teachers between the hunting efficiency of the hunters and the total pool of available hunters.

\section{Results}

\subsection{The Optimal Proportion of Teachers}

Here we derive the optimal proportion of teachers \( \tauopt \) in a population of size \(n\). 
We first compute the equilibrium frequencies of amateur and expert hunters from Eq.~\ref{eq:baseline_model}: 
\begin{align} \label{eq:baseline_equilibrium}
    (n_0^\ast, \;n_1^*) 
    = \frac{(1-\tau)n}{1 +\eta + \nu \tau n} (1, \;\eta + \nu \tau n), 
\end{align}
where we have introduced two rescaled parameters: the self-learning efficiency \( \eta = \lambda / f \) and the education efficiency \( \nu = \mu / f \). 
Substituting this equilibrium into Eq.~\ref{eq:baseline_productivity}, 
we obtain the equilibrium per-capita productivity as a function of teacher proportion and total population size:
\begin{align}
    F(\tau; n) = (1-\tau)\frac{h_0 + h_1 (\eta + \nu \tau n)}{1 + \eta + \nu \tau n}. \label{eq:equilibrium_productivity}
\end{align}
Taking the derivative of productivity \( F \) with respect to \( \tau \) then yields 
\begin{align}
    \tau = \frac{1}{\nu n} 
    \left[ -(1+\eta) \pm \sqrt{\left(1 - \frac{1}{r} \right)(1 + \eta + \nu n)} \right],
    \label{eq:tau_sol}
\end{align} 
where \( r = h_1 / h_0 \) represents the \emph{advantage of expertise} relative to amateur skill (see Supporting Information for a detailed derivation). 
Here we must preclude the possibility that the solution becomes negative, 
which gives rise to 
\begin{align} \label{eq:tauopt_sol}
    \tauopt = 
    \begin{cases}
        0 &\quad n < n_c, \\
        \taup &\quad n > n_c,
    \end{cases}
\end{align}
where \( \taup \) denotes the solution (Eq.~\ref{eq:tau_sol}) with positive sign. As this expression shows, there is a critical population size \( n_c \) required before it is beneficial to dedicate anyone to teaching, given by
\begin{align}
    n_c 
    = \frac{(1 + r \eta)(1 + \eta)}{\nu (r-1)}. \label{eq:n_c}
\end{align}
We note that $n_c>0$ always exists---meaning that teachers are beneficial only if the population size is large enough---even when $\eta =0$, that is, even when expertise requires dedicated teaching and cannot be acquired by self-learning.

\begin{figure}[t]
    \centering
    \includegraphics[width=1\linewidth]{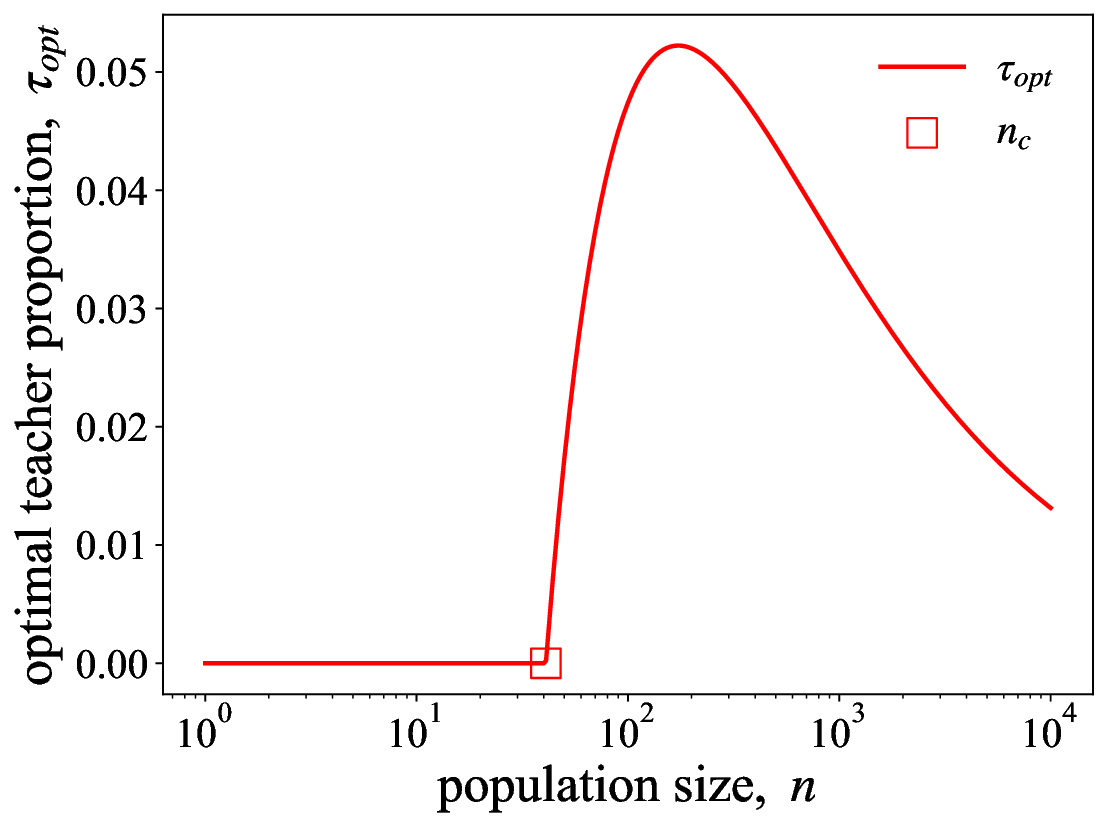}
    \caption{The optimal proportion of teachers \( \tauopt \) as a function of population size \( n \). The square symbol indicates the minimum population size \( n_c \) that is required to justify designating anyone at all to the teaching role. As population size \( n \) increases, the optimal proportion of teachers \( \tauopt \) reaches its maximum and slowly decreases afterwards. Parameters: \( h_0 = 8, h_1 = 10, \lambda = 0.1, \mu = 1, f = 10 \). }
    \label{fig:tauopt_profile}
\end{figure}

\subsection{The Non-monotonic Effect of Population Size on Optimal Teaching}

The optimal proportion of teachers \( \tauopt \) depends on the total population size (see Fig.~\ref{fig:tauopt_profile}). For small populations the optimal proportion of teachers is strictly zero, and it remains zero until the population size reaches the critical value \( n_c \) required to support any teachers. Thereafter, the optimal proportion of teachers increases rapidly, reaches its maximum for an intermediate population size, and then gradually declines at a rate proportional to \( 1/\sqrt{n} \). In other words, there is a non-monotonic relationship between the population size and the optimal proportion of teachers.

There is a simple intuition for the overall behavior of the optimal proportion of teachers \( \tauopt \) as a function of population size \( n \). 
The basic trade-off between efficient hunting achievable through education and the total pool of available hunters implies that some intermediate proportion of teachers leads to the greatest hunting productivity. 
However, the optimal proportion of teachers \( \tauopt \) tends to be substantially pulled toward zero depending on the population size \( n \), as can be seen in Fig.~\ref{fig:tauopt_profile}. 
To understand this, we point out that the per-capita productivity 
\( F \) (see Eq.~\ref{eq:equilibrium_productivity})
is a product of the relative frequency of hunters \( 1 - \tau\) and a weighted average of the two hunting rates \( \left[h_0 + h_1 (\eta + \nu \tau n)\right]/(1 + \eta + \nu \tau n) \), which 
always falls between \( h_0 \) and \( h_1 \) by definition. 
If the population size \( n \) is small, then,
improving the weight-averaged hunting rate 
by increasing the frequency of teachers not only fails to cancel out the consequent labor loss, but it can detract significantly from the overall productivity, which helps explain why there should not be any teachers at all for small \(n\). 
In a larger population, however, increasing the proportion of teachers by only a \emph{slight} amount can even enhance the productivity so much that it not only offsets the resulting labor loss but it elevates the overall productivity to values as large as the expert hunting rate \( h_1 \), implying that the proportion of teachers \( \tau \) should be positive but as small as possible. 
Taken together, these qualitative insights align with the non-monotonic behavior of the optimal teacher profile seen in Fig.~\ref{fig:tauopt_profile}.

\subsection{A Discrete Stochastic Model}

\begin{figure*}[t]
    \centering
    \includegraphics[width=\textwidth]{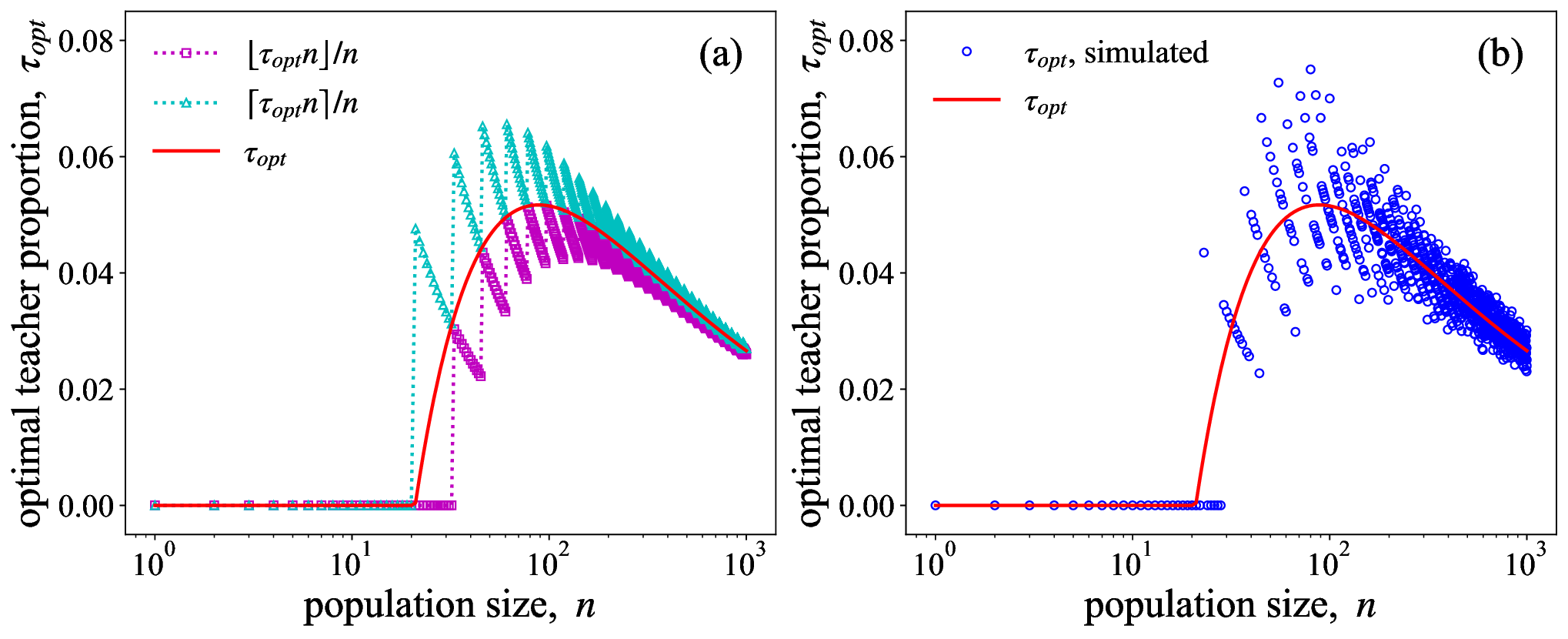}
    \caption{(a) Theoretical predictions for realizable optimal fraction of teachers based on floors (purple squares) and ceilings (light blue triangles) of the continuous model. (b) Numerical results (blue circles) obtained by Monte Carlo simulations of a corresponding discrete state Markov process. The red line in both panels represents the optimum proportion of teachers given by Eq.~\ref{eq:tauopt_sol}. The continuous optimal profile serves as a good predictor of the discrete state Markov model.  When adjusted to ensure that the number of teachers is an integer (Fig.~\ref{fig:gillespie_and_theory}a), its behavior closely matches that  of the discrete stochastic process (Fig.~\ref{fig:gillespie_and_theory}b). 
    Parameters: \( h_0 = 8, h_1 = 10, \lambda = 0.1, \mu = 1, f = 5 \). 
    } 
    \label{fig:gillespie_and_theory}
\end{figure*}

Figure~\ref{fig:tauopt_profile} illustrates the relationship between population size \( n \) and the optimal proportion of teachers \( \tauopt \) under a given set of model parameters. 
However, there is a limitation to the physical interpretation of this continuous profile: the resulting optimal \emph{number} of teachers \( \tauopt n \) might be non-integer, even if \( n \) takes positive integer values. 
In fact, 
the numbers included in the set \( \{ \tauopt (n) n \}_{n=1,2, \ldots} \) are not integers, which implies that the optimal number of teachers, when positive, predicted by our analysis is not physically realizable. This deficiency arises from our treatment of frequencies as real numbers despite the finite population size \(n\).

To resolve this deficiency, we develop a corresponding discrete-state continuous-time Markov process whose mean behavior in the limit of a large population is given by the ODE in Eq.~\ref{eq:baseline_model}. 
Let \( n_0 \) and \( n_1 \) be both nonnegative integers. 
Over a small time interval, we stipulate that the number of amateur (or expert) hunters either increases or decreases by at most one. Since there are three possible events---self-learning, education, and forgetting---the transition \( (n_0, n_1) \to (n_0+1, n_1-1) \) occurs with probability \( f n_1/ p_{tot}\), and the transition \( (n_0, n_1) \to (n_0-1, n_1+1) \) occurs with probability \( (\lambda + \mu n_T)n_0 / p_{tot}\), where \( p_{tot} = f n_1 + (\lambda + \mu n_T)n_0 \) is total rate of either event. 
The time taken for an either event to occur is distributed according to an exponential distribution with mean \( p_{tot} \). 
For each possible (integer) number of teachers, \( n_T \in \{0, 1, \ldots, n\} \), 
we can compute the time-averaged productivity and identify the discrete teacher-to-population ratio \( \tau = n_T/n \) that maximizes the productivity over a sufficiently long time period.

We have implemented Monte Carlo simulations of this Markov process to investigate how the discrete nature of the population affects the optimal proportion of teachers. We can compare these results to a rudimentary theoretical prediction based solely on our continuous model, simply by taking the floor and ceiling---\( \lceil \tauopt n \rceil / n \) and \( \lfloor \tauopt n \rfloor / n \)---to produce physically realizable predictions for the optimal proportion of teachers. As shown in Fig.~\ref{fig:gillespie_and_theory}, these theoretical predictions provide a close match to the results of the discrete Markov model \cite{Gardiner1985}.

\subsection{A Critical Population Size Always Exists}

One of our key findings is that a population must reach a critical size before there is any demand for dedicated teachers. We can prove that this result holds regardless of parameter values (see Sec.~\ref{sec:matmet} for details). 
This argument depends on the following observations about the per-capita productivity: 
(i) \( \partial_{\tau}^2 F (\tau; n) < 0 \) for all \( \tau \in [0,1]\), and 
(ii) \( \partial_{\tau} F(0; n) \) increases linearly with \( n \), 
satisfying \( \partial_{\tau} F(0;0)<0 \) and \( \partial_{\tau} F(0;n) > 0 \) for sufficiently large \( n \) (see Sec.~\ref{sec:matmet}). 
Depending on the population size \( n \), 
there are two possible regimes: \( \partial_{\tau} F(0; n) < 0 \) or \( \partial_{\tau} F(0; n) > 0 \). 
In the first regime, 
\( F \) decreases monotonically from \( F(0;n) > 0 \) to \( F(1;n) = 0 \), 
implying the absence of a peak (or maximum productivity). 
In the second regime, 
given that \( \partial_{\tau} F(1; n)  < 0\) (see Sec.~\ref{sec:matmet}), 
the \emph{intermediate value theorem} guarantees that
there exists a \( \tau \in (0,1) \) such that \( \partial_{\tau} F(\tau; n) = 0 \). 
Since \( \partial_{\tau} F \) is monotonically decreasing with respect to \( \tau \), 
the intersection point is unique, 
demonstrating the presence of a single peak. 
In simple terms, productivity
\( F \) must ``go up'' first and then ``go down'' afterwards, as a function of teacher proportion. 

Because \( \partial_{\tau} F(0; n) \) is monotonically increasing in \( n \), \( \partial_{\tau} F(0; n) = 0 \) yields a unique positive solution \( n_c \). 
This critical population size \( n_c \) 
distinguishes the two distinct regimes mentioned above, 
regardless of other parameter values, 
completing the proof.

\subsection{How the Critical Population Size for Teaching Depends on Parameters}

Partial derivatives of the critical population size \( n_c \) reveal the contribution of each model parameter (see Table~\ref{tab:partials} for a list of partials).
For example, increasing the self-learning rate \( \lambda \) results in an increase in the critical population size \( n_c \), because amateur hunters can learn to hunt more efficiently by themselves. 
On the other hand, 
increasing the education rate \( \mu \) or the benefit of expertise \( r \) each independently decreases the critical population size \( n_c \), because more efficient education or an increased contribution of expertise to productivity both incentivize allocating more individuals as teachers, regardless of other parameter values.
See Sec.~\ref{sec:matmet} for the full expressions of these partial derivatives. 

\begin{table*}[tb]
    \centering
    \renewcommand{\arraystretch}{1.5} 
    
    \begin{tabular}{c l} 
        \hline \hline
        Partial & Comment \\
        \hline
        \( \frac{\partial n_c}{\partial \eta} > 0 \) & Increasing self-learning efficiency \( \eta \) produces larger \( n_c \). \\
        \( \frac{\partial n_c}{\partial \nu} < 0 \) & Increasing education efficiency \( \nu \) produces smaller \( n_c \). \\
        \( \frac{\partial n_c}{\partial r} < 0 \) & Increasing expert hunting rate \( h_1 \) relative to the amateur hunting rate \( h_0 \) produces smaller \( n_c\). \\
        \hline \hline
    \end{tabular}
    \caption{Parameter dependence of the critical population size \( n_c \). See Sec.~\ref{sec:matmet} for full expressions.}
    \label{tab:partials}
\end{table*}

\subsection{Uniform Upper Bound on Teacher Allocation (\( 50 \)\% Rule)}

How large can the optimal proportion of teachers \( \tauopt \) be? 
Is there any scenario in which a majority of the population should be teachers?
To provide full-fledged answers to these questions, 
we first solve 
\begin{align*}
    \frac{\partial \taup}{\partial n} = 0 
\end{align*}
to find the population size \( n^\ast\) at which the hunting productivity attains its highest value, provided that other parameters are fixed. 
This gives
\begin{align}
    n^\ast = \frac{2(1+\eta)\left[(r\eta+1) + \sqrt{r(1+\eta)(r\eta+1)}\right]}{\nu (r-1)} \label{eq:n_star}
\end{align}
(see Supporting Information for a detailed derivation). 
Let \( \tauopt^\ast =  \max_{n > 0} \{\tauopt (n)\} \), 
i.e.,
\( \tauopt^\ast = \taup (n^\ast)\), denote the largest possible value for the optimal proportion of teachers over all possible population sizes. 
Substituting Eq.~\ref{eq:n_star} into 
\( \tauopt \), we obtain  
\begin{align}
    \tauopt^\ast = \frac{r-1}{2\sqrt{r(1+\eta)}\left[\sqrt{1 + r \eta} + \sqrt{r(1 + \eta)} \right]}. 
    \label{eq:tauopt_max}
\end{align}
Because \( 1 + r\eta > 1 + \eta\), we have established the following uniform upper bound on the optimal proportion of teachers: 
\begin{align*}
    \tauopt^\ast 
    < \frac{1}{2}. 
\end{align*}
In other words, 
assigning more than half of a population to teachers can \emph{never} be optimal in this framework, regardless of parameter values.

\begin{figure}[t]
    \centering
    \includegraphics[width=1\linewidth]{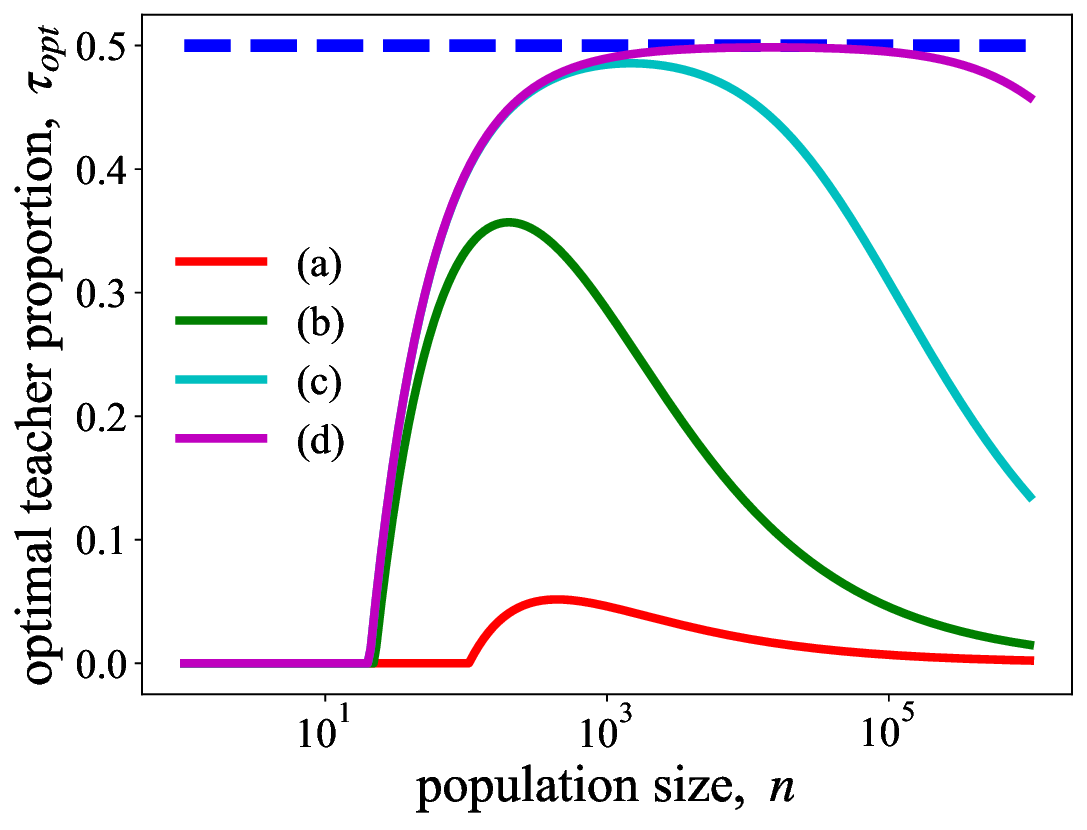}
    \caption{Upper bound on the optimal proportion of teachers (blue dashed line). \( \tauopt^\ast \) approaches the uniform upper-bound of \( 1/2 \) as \( r \to \infty \) and \( \eta \to 0 \), i.e., when expertise dramatically increases productivity but is considerably difficult to acquire. In addition, the maximal proportion of teachers $\tau=1/2$ can only be reached for populations of an intermediate size. 
    Parameters: \( h_0 = 8, \lambda = 0.1, \mu = 0.2 \). 
    (a) \( f = 5, h_1 = 10 \).  
    (b) \( f = 50, h_1 = 100 \).  
    (c) \( f = 5000, h_1 = 10000 \).  
    (d) \( f = 500000, h_1 = 1000000 \). 
    }
    \label{fig:uniform_upper_bound}
\end{figure}

There is a simple explanation for this 50\% rule on maximum teacher allocation. 
Intuitively, the benefit of teachers will be greatest when the skill is very difficult to master and it substantially boosts productivity once acquired---i.e., when
\( \eta, \nu \ll 1 \) and \( r \gg 1 \), 
which we call the \emph{expertise-limited} regime. 
In this regime, to achieve the highest possible hunting productivity 
we must maximize the number of expert hunters (because amateur hunting contributes a negligible amount to total productivity) even in the face of potential labor loss due to teacher allocation.
Naively, since most hunters will remain amateur in this regime of difficult learning, 
i.e., \( n_0 \approx (1-\tau) n \), 
and because \( n_T = \tau n \) by definition, 
the number of expert hunters will be maximized when approximately half of the population engages in teaching 
(note, however, that we have effectively ignored 
the fact that increasing the proportion of teachers \( \tau \) can \emph{decrease} the number of expert hunters \( \mu n_T n_0\) that are produced per unit time, depending on the population size \( n \)
[see Eq.~\ref{eq:baseline_equilibrium}]). 
More precisely, assuming that \( r \gg 1 \) and \( \eta \ll \nu \tau n \), 
the per-capita productivity behaves roughly as 
\( F \propto (1-\tau) \nu \tau n / (1 + \nu \tau n) \). 
If the term \( \nu \tau n \) is sufficiently smaller than \( 1 \) (which will be verified post hoc), 
we can asymptotically expand productivity as 
\( F \propto (1-\tau) \nu \tau n + O((\nu \tau n)^2) \), 
which is maximized at \( \tau \approx 1/2 \). 
The approximation above is valid only if the condition \( \eta \ll \nu \tau n \ll 1 \) is met, 
which means that the optimal proportion of teachers, \( \tauopt \), will be \( \approx  1/2 \) only when the population size \( n \) is \emph{moderately} large, as seen in Fig.~\ref{fig:uniform_upper_bound}, in which case the argument becomes self-consistent.

\subsection{Asymptotic Upper Bound on Teacher Allocation (\( 25 \)\% Rule)}

\begin{figure}[t]
    \centering
    \includegraphics[width=1\linewidth]{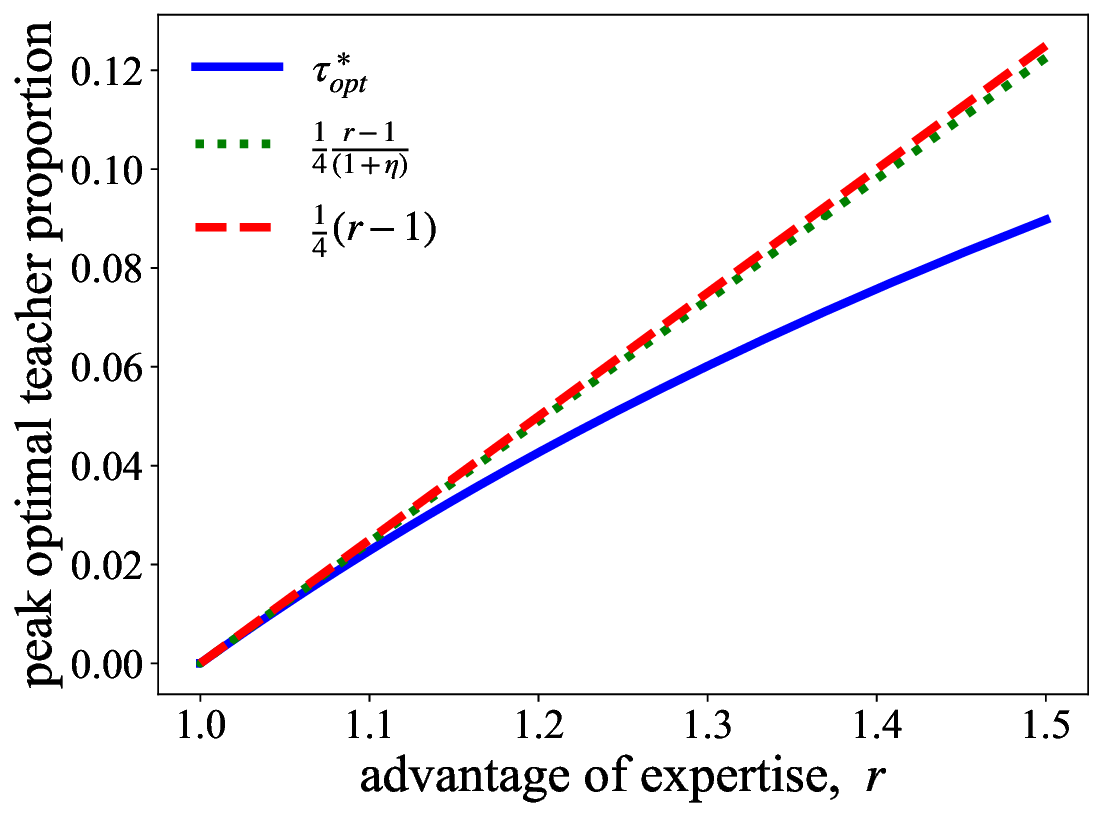}
    \caption{The \( 25 \)\% rule of thumb (asymptotic tight upper bound on \( \tauopt^\ast \)). If there is little advantage to expertise and self-learning is difficult, then peak educational demand---which occurs at \( n = n^\ast \)---scales as \( 25 \)\% of the relative advantage of expertise,  \( r-1 \). Parameters: \( h_0 = 8, \lambda = 0.1, \mu = 0.2, f = 5 \). 
    }
    \label{fig:tight_upper_bound}
\end{figure}

The maximal optimal proportion of teachers \( \tauopt^\ast \) decreases to zero as the advantage of expertise disappears, that is, as \( r \to 1 \). 
This is because there is no need for teachers when expertise does not contribute to increasing the hunting productivity. 
However, the asymptotic behavior of Eq.~\ref{eq:tauopt_max} in this limit provides a useful and simple estimate of the \emph{peak} educational demand. 

To derive the asymptotic behavior of the maximal proportion of teachers \( \tauopt^\ast \) in the limit of low educational improvement (\( r \to 1 \)), 
let us consider the following function 
\begin{align*}
    \mathcal{G}(x) := \frac{x-1}{\sqrt{x}\left(1+\sqrt{x}\right)}, 
    \quad x > 1, 
\end{align*}
which can be approximated as
\begin{align*}
    \mathcal{G}(x) 
    \approx \mathcal{G}'(1)(x-1) =\frac{1}{2} (x-1)
\end{align*}
around \( x = 1 \), up to first order.  
Piecing together the above and previous results, we find that 
the highest optimal proportion of teachers \( \tauopt^\ast \) has a tight upper-bound 
\begin{align*}
    \tauopt^\ast 
    \lesssim \frac{1}{4(1+\eta)}(r-1) 
\end{align*}
in the limit \( r \to 1 \) (see Supporting Information for details). 
Note that this tight bound is independent of the education rate \( \mu \). 
Moreover, when expertise brings little advantage and the skill cannot be easily self-learned, then the peak demand for teachers is approximately one \emph{quarter} of the advantage of expertise. This asymptotic bound on the optimal proportion of teachers, in terms of the advantage of expertise, provides some practical guidance for situations when expertise has limited value. 

The two bounds we have derived on maximal teacher allocation are complementary to one another.
The \emph{uniform} upper bound \( \tau=1/2 \) on the proportion of teachers---which is \emph{independent} of all other parameters---can only be achieved when expertise offers immense advantage (Fig.~\ref{fig:uniform_upper_bound}) and population size is intermediate; 
on the other hand, 
the \emph{asymptotic} tight upper bound on \( \tau \) is valid only in the regime where the relative advantage of expertise \( r - 1 \) is small (Fig.~\ref{fig:tight_upper_bound}). Moreover, the asymptotic bound depends upon \( r \) because it describes how fast \( \tauopt^\ast \) decreases as the advantage of expertise decreases (i.e., \( r \to 1 \)).

\subsection{Six Extended Models}

Next we analyze a variety of alternative model formulations to test the robustness of our conclusions. 
We focus on the optimal teacher-to-population ratio \( \tauopt \), and we ask whether the critical population size required for teachers to be beneficial is a general phenomenon. To this end, we consider the following six model variations and extensions (see Table~\ref{tab:five_extensions} and Supporting Information for full mathematical descriptions): 
(a) \emph{part-time hunting} (teachers can hunt a little bit \emph{instead} of teaching); 
(b) \emph{eager beavers} (teachers can hunt a little bit \emph{in addition} to teaching);  
(c) \emph{part-time teaching} (expert hunters teach, i.e., there are no dedicated teachers);  
(d) \emph{on-the-job training} (teaching involves a little bit of hunting); 
(e) \emph{finite teaching capacity} (teachers have a limited potential teaching capacity); and (f) \emph{three-stage model} (three levels of expertise).

\begin{table*}[t]
    \centering
    \renewcommand{\arraystretch}{2.0} 
    \begin{tabular}{l m{0.5\textwidth}}  
        \hline \hline
        Model & Dynamical equations, productivity function \\
        \hline

        Part-time hunting & 
        $ 
        \begin{array}{rl} &
            \begin{cases}
                \dot n_0 = f n_1 - \left[\lambda + \mu (1-\gamma) n_T \right] n_0 \\
                \dot n_1 = \left[\lambda + \mu (1-\gamma) n_T \right] - f n_1 
            \end{cases} \\
            &F = \frac{n_0 h_0 + n_1 h_1 + \gamma n_T  h_T}{n}, h_0 < h_T < h_1. 
        \end{array}
        $ \\
        \hline
        Eager beavers & 
        $
        \begin{array}{rl} &
            \begin{cases}
                \dot{n}_0 = f n_1 -  (\lambda + \mu n_T)n_0, \\
                \dot{n}_1 = (\lambda + \mu n_T) n_0 - f n_1, \\
            \end{cases} \\
            &F = \frac{n_0 h_0 + n_1 h_1 + n_T h_T}{n}, h_0 < h_T < h_1.
        \end{array}
        $ \\
        \hline
        Part-time teaching &  
        $
        \begin{array}{rl}&
            \begin{cases}
                \dot{n}_0 = f\,n_1 - \left(\lambda + \mu \tilde{\tau} n_1\right)\,n_0, \\
                \dot{n}_1 = \left(\lambda + \mu \tilde{\tau}\,n_1\right) n_0 - f n_1, 
            \end{cases} \\
            &F = \frac{n_0 h_0 + (1-\tilde\tau) n_1 h_1}{n}, n = n_0 + n_1.
        \end{array}
        $ \\
        \hline
        On-the-job training &  
        $
        \begin{array}{rl}
            F = \frac{n_0 h_0 + n_1 h_1 + \lambda_1 n_T n_0 h_T}{n}, h_T < h_0 < h_1. \\
        \end{array}
        $ \\ 
        \hline
        Finite teaching capacity & 
        $
        \begin{array}{rl}
            \begin{cases}
                \dot{n}_0 = f n_1 - \left(\lambda + \mu \frac{K}{n + K} n_T \right) n_0, \\
                \dot{n}_1 = \left(\lambda + \mu \frac{K}{n + K} n_T\right) n_0 - f n_1. 
            \end{cases}
        \end{array}
        $ \\
        \hline
        Three-stage model &  
        $
        \begin{array}{rl}
            &
            \begin{dcases}
                \dot{n}_0 = f\,n_1 - (\lambda_0 + \mu_0 \alpha n_T) n_0, \\
                \dot{n}_1 = (\lambda_0 + \mu_0 \alpha n_T) n_0 - f n_1 + g n_2 
                 - \left[\lambda_1 + \mu_1 (1-\alpha) n_T\right] n_1, \\
                \dot{n}_2 = \left[\lambda_1 + \mu_1 (1-\alpha) n_T\right] n_1 - g n_2, 
            \end{dcases} \\
            &
            F = \frac{h_0 n_0 + h_1 n_1 + h_2 n_2}{n}, h_0 < h_1 < h_2.
        \end{array}
        $ \\
        \hline \hline
    \end{tabular}
    \label{tab:five_extensions}
    \caption{Six variations or extension of the baseline hunter-teacher model. The last two models are qualitatively different from the others. All six models exhibit similar qualitative phenomena, including a minimum population size required before teacher allocation is beneficial.}
\end{table*}

The first four model variations---\emph{part-time hunting}, \emph{eager beavers}, \emph{part-time teaching}, and \emph{on-the-job training} models---are introduced to explore the robustness of our results. 
Despite various structural differences, we find that
in these cases there is still a critical population size \( n_c \) required for teaching to be beneficial, and they all exhibit a similar overall behavior of the optimal proportion of teachers \( \tauopt \) as a function of population size \( n \) 
(see Supporting Information for more details). 
It is interesting to note that even in the part-time teaching model, where experts can choose how much effort they devote to teaching compared to hunting, there is a regime where the optimal productivity is achieved by expert hunters dedicating almost all their time to teaching amateurs---i.e., a \emph{dedicated} teacher class emerges in this regime, even if not imposed a priori (see Supporting Information for details).

\begin{figure}[t]
    \centering
    \includegraphics[width=1\linewidth]{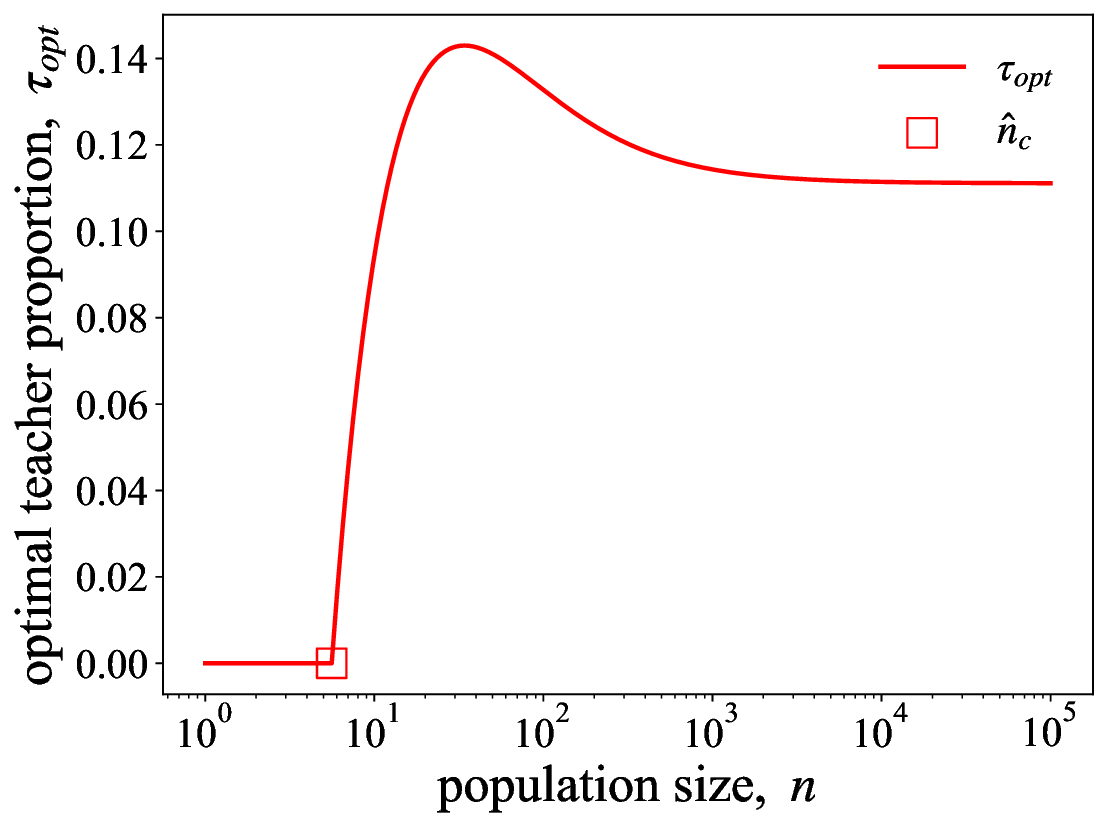}
    \caption{The optimal proportion of teachers \( \tauopt \) for a population of size \( n \) in the finite-capacity model. The optimal proportion does not vanish for large \( n \) but converges to some positive value, which is not the case in the baseline model (cf.~Fig.~\ref{fig:tauopt_profile}). Parameters: \( h_0=5, h_1=10, \lambda_0=0.1, \lambda_1=1, f = 5, K = 100 \).
    }
    \label{fig:finite_capacity_model_optimal_profile}
\end{figure}

The fifth model variation---\emph{finite teaching capacity} model---is qualitatively different from our baseline model, and it serves a different purpose. 
In our baseline model (Eq.~\ref{eq:baseline_model}),  
we assumed that teaching occurs in a \emph{density-dependent} manner, so that a teacher can reach an arbitrarily large number of students as the population size increases.
But this assumption is almost surely violated in reality, due to spatial and temporal constraints on the process of teaching and learning. In a more realistic model, the total rate of teaching in a large population does not depend on the absolute number of teachers but rather on their relative frequency in the population.  To incorporate this effect of finite teaching capacity, 
we now suppose that the education term, \( \mu n_T n_0 \), remains the same for small \( n \) but is divided by \( n \) for large \( n \). 
One way to achieve this---without losing continuity and smoothness---is multiply the term by a factor of \( \chi(n) = K / (n + K) \), where \( K \) represents the \emph{50\% saturation point} of teaching, i.e., the population size at which the teaching capacity drops to half of its maximum. 
Figure~\ref{fig:finite_capacity_model_optimal_profile} demonstrates how the optimal fraction of teachers \( \tauopt \) changes with population size \( n \) in the  model with finite teaching capacity. 
Although the overall shape of this profile is similar to Fig.~\ref{fig:tauopt_profile}, the optimal proportion of teachers \( \tauopt \) does not vanish as the population size grows but saturates at some finite positive value (see also Supporting Information).

The finite-capacity model produces a more nuanced criterion for whether teaching is beneficial or not:
teachers are necessary only if the \( 50 \)\% saturation size \( K \) exceeds the critical population size \( n_c \) in the baseline model, whereas if \( K \) is smaller than \( n_c \) then having no dedicated teachers is optimal regardless of population size. 
These results for the finite-capacity model are summarized as follows: 
If \( K > n_c \), then 
\begin{align*}
    \tauopt &=
    \begin{cases}
    \hat \tau_+ &\quad n > \hat n_c, \\
    0 &\quad n < \hat n_c, 
    \end{cases}
\end{align*}
where 
\begin{align*}
    \hat n_c &= \frac{K n_c}{K-n_c},  \\
    \hat\tau_{+} 
    &= 
    \frac{ - (1+\eta) + \sqrt{(1 - 1/r)(1+\eta +\nu \chi(n)n)}}{\nu \chi(n) n} 
\end{align*}
If \( K < n_c \), on the other hand, we have \( \tauopt = 0 \) for all \( n \). 
We refer the reader to Supporting Information for more details. 
To conclude, 
whether or not there is any demand for teachers is contingent on 
the potential teaching capacity \( K \) as well as overall population size \( n \).

\subsection{A Model with Three Levels of Expertise} 

\begin{figure*}[t]
    \centering
    \includegraphics[width=1\textwidth]{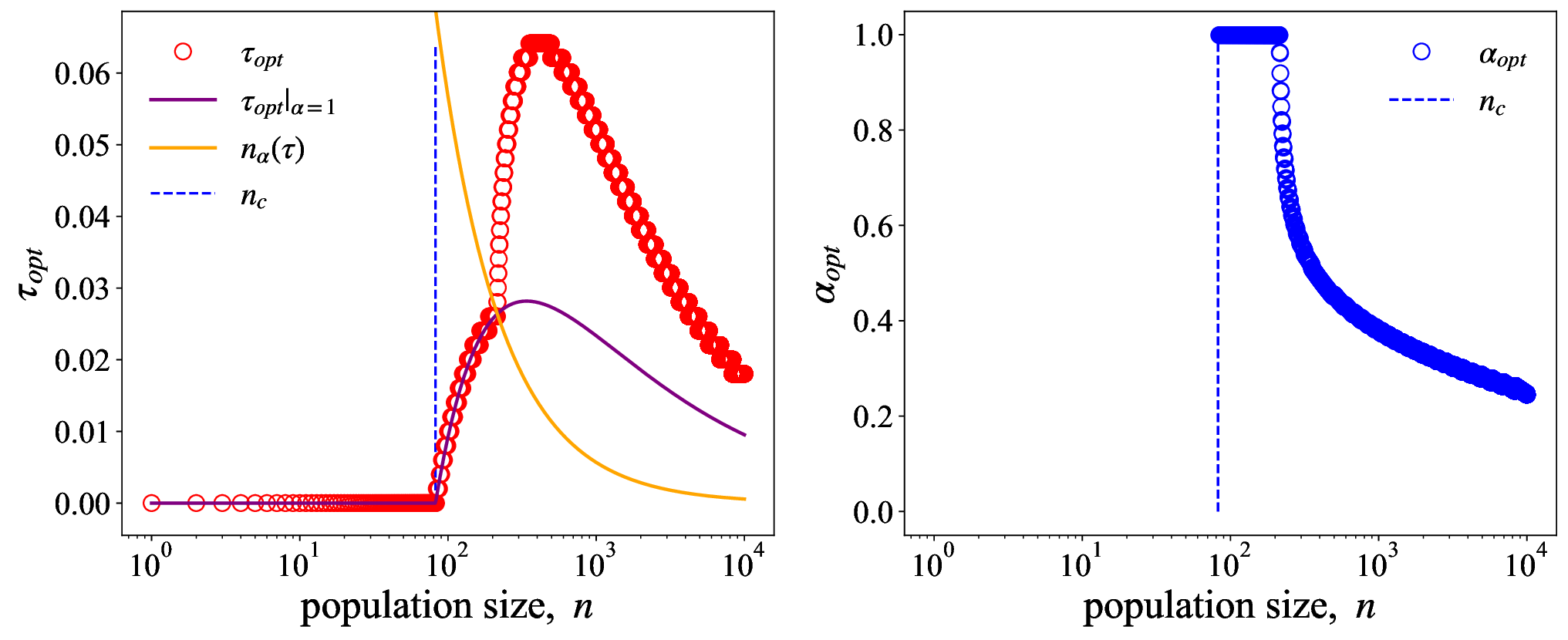}
    \caption{The optimal proportion of teachers \( \tauopt \) within a population (red line, left panel) and the optimal proportion of teachers who educate the least skilled hunters (\( \alphaopt \), blue line, right panel) for different population sizes \( n \), derived by numerically maximizing productivity of the three-stage model. The optimal proportion of teachers exhibits two transitions, each of which results in a sudden increase. The optimal profile of \( \alpha \) implies that all teachers should be devoted to educating the least skilled group of hunters \( A_0 \) for some intermediate range of \( n \); and in larger populations an increasing fraction of teachers should teach the second least skilled hunters \( A_1 \).  Parameters: \( h_0 = 8, h_1 = 9, h_2 = 10, \lambda_0 = 0.2, \lambda_1 = 0.1, \mu_0 = \mu_1 = 1, f = 10, g = 15 \). 
    }
    \label{fig:three_stage_model_optimal_profile}
\end{figure*}

The final extension we explore is a \emph{three-stage} model, 
in which expertise develops through three different levels
 \( A_0 \to A_1 \to A_2 \), which have associated hunting rates, \( h_0 \), \( h_1 \), and \( h_2 >h_1 >h_0\)  corresponding to, e.g., amateur, intermediate, and expert classes. 
Since there are two sequential but distinct processes of skill acquisition, 
teachers can be assigned to facilitate either one of these transitions. 
Let \( \alpha \) (\( 0 \leq \alpha \leq 1 \)) be the ratio---relative to the total number of teachers---of those who are dedicated to teaching the beginner's class \( A_0 \) 
(see Table~\ref{tab:five_extensions} for a full definition of the model). 
Aside from the fact that there are now two 
degrees of freedom, \( \tau \) and \( \alpha \), 
this new scenario is otherwise similar to the baseline model. 
However, this model offers a more nuanced understanding of the evolution of teachers because it addresses not only the optimal number of teachers but also what level of students they should teach.

Figure~\ref{fig:three_stage_model_optimal_profile} illustrates how the optimal proportion of teachers, \( \tauopt \), and the fraction of those who should be dedicated to teaching the \emph{amateur} class of hunters \( A_0 \), \( \alphaopt \), change with population size \( n \). 
Notably, 
\( \tauopt \) now has two critical population sizes, \( n_c^\ast \) and \( n_{\alpha}^\ast \), in contrast to Fig.~\ref{fig:tauopt_profile}. 
The first critical population size \( n_c^\ast \) occurs when the need for any teachers emerges for the first time. 
The second critical population size \( n_\alpha^\ast \) occurs when the optimal proportion of teachers \( \alphaopt \) assigned to the second stage of skill acquisition becomes nonzero; in other words, 
there is a regime (i.e., \( n \in [n_c^\ast, n_\alpha^\ast] \)) where 
\emph{all} teachers should educate the \emph{least} skilled hunters \( A_0 \) before anyone is allocated to educate the intermediate class of hunters \( A_1 \) in response to a further population increase. 
In short, 
we find a \emph{two-tier} transition in the optimal proportion of teachers, which provides a more refined understanding of the evolution of teachers, including how teaching efforts should be allocated across learners of different levels of expertise.

To analyze this model thoroughly, 
we must solve 
\( \partial_\tau F = 0, \; \partial_\alpha F=0 \). 
However, due to their algebraic complexity, we cannot in general find closed-form solutions. 
Instead, we provide an ad hoc but effective approach to identify the two critical population sizes \( n_c^\ast \) and \( n_\alpha^\ast \), as indicated in Fig.~\ref{fig:three_stage_model_optimal_profile}, under suitable conditions. 
Specifically, we impose that a nonzero \( \tauopt \) emerges continuously from \( \tauopt = 0 \). 
In Supporting Information we derive 
\begin{align}
    n_c^\ast = \frac{(1 + r_1 \eta_0 + r_2 \eta_0 \eta_1) (1 + \eta_0 + \eta_0 \eta_1)}{\nu_0 [(r_1 - 1) + (r_2 - 1)\eta_1]} 
    \label{eq:n_c_three_stage_model}
\end{align}
via a suitable change of parameters (\( \eta_0 = \lambda_0 / f, \eta_1 = \lambda_1 / g, \nu_0 = \mu_0 / f, \nu_1 = \mu_1 / g, r_1 = h_1/h_0, r_2 = h_2/h_0 \)). Comparing Eq.~\ref{eq:n_c_three_stage_model} to Eq.~\ref{eq:n_c} shows how the extra layer of expertise affects the critical population size required for a dedicated teacher class to emerge. 
We also explain in Supporting Information why \( n_\alpha^\ast \) emerges as the intersection of the two curves \( \left.\tauopt \right|_{\alpha=1} \) and \( n_\alpha (\tau) \propto 1/\tau \). 
The restricted function \( \left.\tauopt \right|_{\alpha=1} \) represents the optimal proportion of teachers, given that teachers educate only the \emph{least} skilled hunters \( A_0 \) (but not intermediate hunters \( A_1 \)), which is based on the assumption that a nonzero \( \tauopt \) arises \emph{continuously} from \( 0 \). Similarly, \( n_\alpha (\tau) \) is derived by solving \( \partial_{\alpha} F (\tau,1;n)  = 0 \), which arises from the assumption that the optimal fraction of beginners' teachers \( \alphaopt \) drops \emph{continuously} from \( 1 \) as the population grows beyond \( n_c^\ast \).  

We conjecture that in this \emph{three-stage model} the uniform upper bound on the optimal proportion of teachers is \( 2/3 \), based on our previous intuitive argument in the two-stage model; and, likewise, that the ``\( M \)-stage model'' has an upper bound of \( M /(M-1) \) on the optimal proportion of teachers, which can be achieved in an expertise-limited regime.

\section{Discussion}

One of our key results is that a population must exceed a critical size before it is beneficial to allocate anyone as a dedicated teacher. Subsequently, the peak demand for teachers is achieved at some intermediate population size, and the peak demand never surpasses one half of the population. 
Several extensions of our baseline model confirm the typical behavior of optimal teacher allocation as a function of population size and model parameters. 
When the idealized assumption of density-dependent teaching is relaxed, the optimal proportion of teachers converges to a finite positive value in the limit of large population, which aligns with our intuition. 
An extended model that incorporates multiple levels of expertise illuminates a more refined understanding of optimal allocation of teachers---that is, 
the \emph{layered}, \emph{nonlinear}, and \emph{emergent} complexity in the optimal distribution of teaching efforts.

These results have several implications for optimal organization of a society.
One simple and robust finding is that a society will rarely benefit from a large proportion of dedicated teachers. Even the uniform upper bound of one-half teachers is achievable only in special and extreme cases, when the population size is intermediate and skill is both indispensable and nearly impossible to attain without tutelage.
More generally, our analysis provides guidance on how the proportion of dedicated teachers should be allocated, in terms of the value of expertise, educational efficacy, the rate of self-learning, as well as the population size.
While many distinct factors figure into this equation,
it is critical to recognize that nobody should be dedicated to teaching when the group size is very small, even when the skill cannot be self-learned. More generally, when there are multiple levels of expertise, a population should emphasize earlier stages of learning more than later, and especially so in smaller groups or organizations. These results suggest, for example,
that a small postgraduate institute might be better off focusing solely on teaching master's students rather than offering any formal doctoral programs, whereas a large university might benefit from placing some emphasis on educating doctoral students.

Although our modeling framework is not directly based on prior literature, we can compare our results to prior theoretical and empirical studies of population stratification and productivity across a range of contexts, such as multicellular organisms and microbial cooperation \cite{Pichugin2017, Cornforth2012}. 
In the context of chemical kinetics, for example,
Tannenbaum \cite{Tannenbaum2007} discusses when ``division of labor'' is favored, addressing whether catalysts engaged in one of a series of sequential subtasks outperform simultaneous single-task implementation in final chemical production. 
One of their main conclusions is that a differentiated strategy outperforms an undifferentiated one only if the number of catalysts implementing the task exceeds a threshold. 
This is reminiscent of how our model predicts a critical population size \( n_c \) required for dedicated teachers to emerge. 
However, there are a number of structural differences between these two models. 
The hunter-teacher population aims to maximum (per-capita) productivity given a fixed number of individuals, which is achieved by the optimal allocation of teachers we have identified. 
The compartment model in \cite{Tannenbaum2007}, on the other hand, seeks to maximize the final chemical production (regardless of intermediates) given a fixed number of catalysts, with constant replenishment of the convertible chemical resource.

Prior studies on teaching in animals and humans have highlighted a tension between social and individual learning, which controls the balance of information spread versus novelty acquisition in a population, and therefore shapes the evolution of cumulative culture \cite{Aoki2005, Fogarty2011, Maisonneuve2025}. 
Our study, by contrast, has focused on a different tension, between the expertise of a labor force versus the size of the labor force allocated to production.

The critical population size \( n_c \) for the emergence of a teacher class may provide a framework to understand the so-called \emph{reproductive threshold} that some organisms face as they develop. 
Many species of ants and honeybees, for example, do not produce queens or males until they reach a sufficiently large number of workers, i.e., a minimum reproductive size \cite{Powell2021, Smith2014}. 
Although little is known about its biological origins, our model, when properly modified to link productivity \( F \) to reproductive output, might help explain the adaptive advantage of suppressing reproduction in populations below a threshold size. 

Although only coarse comparisons are reasonable, it is instructive to compare the optimal proportion of teachers in our model to the empirical frequency of dedicated teaching in natural populations.
For individual cells, on one end of the spectrum, chaperone proteins comprise \( 0.3 \)--\( 10 \)\% of the proteome \cite{Rebeaud2021, Shemesh2021}. 
At the other end of the spectrum, in human societies, dedicated teachers comprise a small portion of the population: in the U.S., roughly \( 2.5 \)\% of the workforce (or 1\% of the population) are elementary or secondary teachers \cite{Irwin2024}. 
The teacher-to-population ratio typically ranges from one to a few percent among OECD countries \cite{OECD2024, WorldBank2022}. 
Alternatively, we can consider what proportion of a population is post-reproductive, under the theory that post-reproductive females provide an adaptive benefit by educating future generations \cite{Foster2012,Cant2008, Lahdenpera2014}.
Grandparents, those aged \( 65 \) or older, comprise approximately \( 17 \)\% of the U.S. population. This number is larger in Japan, where the elderly account for nearly  \( 28 \)\% of the population \cite{Caplan2023}.
By contrast, in killer whales post-reproductive females typically comprise \( 11 \)\% of the population \cite{Olesiuk2005}, 
whereas in elephants \( 13 \)\% of adult females are post-reproductive \cite{Lahdenpera2014}. 
Likewise, the proportion of the post-reproductive lifespan relative to an entire lifetime varies widely across mammalian species, from roughly one-half in humans, to one-quarter in whales, to one-eighth in elephants \cite{Lahdenpera2014, Ellis2018}. 
Although currently not addressed by our model, the sources of cultural and cross-specific variation in these statistics is an intriguing topic for future work.

The extended version of our model that allows for three levels of expertise provides new perspectives on the relationship between population size, division of labor, and \emph{complexity}. 
Complexity could arise as an outcome of increasing population size that drives production division of labor within a population. 
According to the \emph{the size--complexity hypothesis} \cite{Feinman2011},
organizational complexity is expected to correlate with population size.
Organizational complexity has been operationalized as 
the number of political types \cite{Johnson1982} or administrative levels \cite{Feinman1984} within a society. The basic results from our model---in particular, the sequential unfolding of critical population sizes that coincide with increasing complexity within the teacher class---aligns well with this general hypothesis. 
Non-human species also seem to follow this pattern in both reproductive and non-reproductive divisions of labor \cite{Bell1997, Bonner2004}: 
a recent study has found consistent evidence in ant societies that larger colonies  promote the evolution of a more diverse set of worker castes \cite{BellRoberts2024, FergusonGow2014}.
The authors reason that species with larger colonies are at a ``lower risk of losing essential worker functions if workers are lost,'' which expresses a tension similar to the trade-off between labor force and per-capita productivity that underlies our results.

Despite the implications described above, our model is certainly not without limitations. First, we have assumed that teaching occurs in a well-mixed population, described by a mass-action term. In reality, however, spatial and temporal constraints may limit the reach of any one teacher. This could undermine the effect of education and increase the critical population size \( n_c \) required for dedicated teaching to emerge. Our model also assumes that learning and teaching occur much faster than population growth. Although realistic for many scenarios, there are certainly conditions in which a population can grow faster than skill acquisition. The consequences of rapid population growth on teaching demand remain an open question.

Perhaps the biggest limitation of our study---and an important direction for future work---is that our analysis is not truly evolutionary. We have studied what allocation of teaching effort is optimal for per-capita production, without studying what individual-level selective forces might, or might not, achieve the outcome that is optimal for the population as a whole. 
One evolutionary pathway towards the emergence of teaching is inclusive fitness.
Indeed, empirical findings show that the prolonged lifespan of post-fertile females increases inclusive fitness through their contribution to the life expectancy and reproductive output of their grandchildren \cite{Lahdenpera2004, Sear2008}. 
Therefore, inclusive fitness effects can help to explain the emergence and long-term dynamics of \emph{vertical teaching} effort---or vertical transmission of skill, from parents to children \cite{Mullon2017}. But teaching is not always vertical. 
\emph{Oblique teaching}, between unrelated individuals \cite{Mullon2017}, is also common among humans, and it is harder to explain by individual fitness alone. One type of mechanism, assumed a priori in our analysis, is an incentive structure that ensures per-capita productivity from acquired skill will be shared equally among all individuals, in which case there will be mutual benefits for oblique teaching. This idealized assumption makes sense for certain types of skilled activities---such as group hunting---but it is violated in other settings. Lacking any mechanism to coordinate incentives across an entire population, dedicated oblique teaching might be restricted to smaller subgroups of the population that benefit equally from skill acquisition within the subgroup. Whether such subgroups will nonetheless be large enough to surpass $n_c$ remains an open question. Likewise, the emergence of institutions or social norms that facilitate the distribution of per-capita production and thereby incentivize oblique teaching in larger populations remains a worthy topic for future research.

Humans are set apart from other species by their extraordinary tendency to cooperate. They teach not because they cannot do. Rather, they teach because of the indirect benefits they accrue through the structure of a cooperative society. And yet, even when society is structured to share the rewards of skillful productivity, dedicated teaching will not arise when populations are too small, and it will never be a majority profession.

\section{\label{sec:matmet} Materials and Methods}

Here we provide a complete, self-contained proof of the existence of the critical population size \( n_c \) in our baseline model. We also present complete expressions for the partials of \( n_c \) that are discussed in Table.~\ref{tab:partials}. 
In Supporting Information we analyze our baseline model in more detail, including a derivation of \( \tauopt \), the uniform upper bound, and the tight upper-bound, as well as our results on all extended models and a full analysis of the model with three levels of expertise.

\subsection{Preliminaries}

We prove that productivity \( F \) in Eq.~\ref{eq:equilibrium_productivity} is a concave function of the proportion of teachers \( \tau \). 
Applying the chain rule twice, we find that 
\begin{align}
    \frac{\partial F}{\partial \tau} 
    &= - W + (1-\tau) \frac{\partial W}{\partial \tau}  
    \label{eq:F_single_prime}  \\
    \frac{\partial^2 F}{\partial \tau^2} 
    &= - 2 \frac{\partial W}{\partial \tau} + (1-\tau) \frac{\partial^2 W}{\partial \tau^2} 
    \label{eq:F_double_prime}
\end{align}
First, we note that \( F(\tau;n) \) is a product of the two opposing factors, \( 1 - \tau \) and \( W(\tau;n) = [h_0 + h_1 (\eta + \nu \tau n)]/(1 + \eta + \nu \tau n) \), that respectively represent the labor loss due to the presence of dedicated teachers and the weight-averaged hunting rate that increases with the proportion of teachers \( \tau \). 
More precisely, 
given a function \( G(x,y) \) of the following form 
\begin{align*}
    G(x, y)
    = \frac{h_0 x + h_1 y}{x + y}, 
\end{align*}
the weight-averaged hunting rate is computed as \( W = G(w_0, w_1) \), with the corresponding weights given by \( w_0 = 1 \) and \( w_1 = \eta + \nu \tau n \). 
Because \( h_0 < h_1 \) by definition, \( G(x,y) \) increases monotonically with \( y \). 
Therefore, 
\begin{align*}
    \frac{\partial G}{\partial \tau}(w_0,w_1(\tau)) 
    = \frac{\partial G}{\partial y} \frac{d w_1}{dt} > 0, 
\end{align*}
implying that \( W \) is monotonically increasing in \( \tau \). 
Moreover, 
\( G(x,y) \) is a concave function of \( y \) (or alternatively, \( G(x,y) \) is a convex function of \( x \)) 
because 
\begin{align*}
    \frac{\partial^2 G}{\partial y^2}
    = - \frac{2x(h_1-h_0)}{(x+y)^3} < 0. 
\end{align*}
Importantly, the composition of a concave function with an affine function is also concave (i.e., it preserves concavity; see below for a proof). 
Therefore, we have 
\begin{align*}
    \frac{\partial^2 W}{\partial \tau^2} < 0, 
\end{align*}
which, taken together with Eq.~\ref{eq:F_double_prime}, implies that 
\begin{align*}
    \frac{\partial^2 F}{\partial \tau^2} < 0, 
\end{align*}
establishing that \( F \) is concave in \( \tau \), regardless of other parameter values.

\subsection{Proof of the existence of \( n_c \) in the baseline model}

We previously showed that 
\( \partial_{\tau}^2 F (\tau; n) < 0 \) for all \( \tau \in [0,1]\). 
A direct calculation shows that 
\begin{align*}
    \frac{\partial F}{\partial \tau}(0;n) 
    = \frac{\nu (h_1 - h_0) n - (h_0 + h_1 \eta)(1+\eta)}{(1 + \eta )^2},
\end{align*}
which guarantees that 
\( \partial_{\tau} F(0;0)<0 \) and \( \partial_{\tau} F(0;n) > 0 \) for sufficiently large \( n \). 
Depending on the value of \( n \), therefore, \( \partial_{\tau} F(0; n) \) can be either positive or negative. 
If positive, \( F \) decreases monotonically from \( F(0;n) > 0 \) to \( F(1;n) = 0 \), 
implying the \emph{absence} of a peak. 
If negative, on the other hand, 
given that, by Eq.~\ref{eq:F_single_prime}, 
\begin{align*}
    \frac{\partial F}{\partial \tau}(1;n) = - W(1;n) < 0, 
\end{align*}
the \emph{intermediate value theorem} ensures that
there is a \( \tau \in (0,1) \) 
such that \( \partial_{\tau} F(\tau; n) = 0 \). 
Since \( \partial_{\tau} F \) is monotonically decreasing with respect to \( \tau \) (i.e., \( F \) is concave), 
the intersection point is unique, 
demonstrating the \emph{presence} of a single peak. 
Because \( \partial_{\tau} F(0; n) \) is monotonically increasing in \( n \), \( \partial_{\tau} F(0; n) = 0 \) yields a unique positive solution \( n_c \). 
This critical population size \( n_c \) 
distinguishes the two distinct regimes mentioned above, 
regardless of other parameter values, 
completing the proof.

\subsection{Concavity Preservation}

Consider a composition \( Z(t) \coloneqq W(y(t)) \) of a concave function \( W(y) \) and a linear function \( y(t) \coloneqq a t + b\). 
We also define \( y_i (t) = a t_i + b \) (\( i = 1, 2 \)). 
Then, for an arbitrary \( \theta \in [0,1] \), it follows that 
\begin{align*}
y((1-\theta)t_1 + \theta t_2) 
&= a[(1-\theta)t_1 + \theta t_2] + b \\
&= (1-\theta)(at_1+b) + \theta (at_2 + b) \\
&= (1-\theta)y_1(t) + \theta y_2(t).
\end{align*}
Then, given that \( W(y) \) is concave in its independent variable \( y \), we establish
\begin{align*}
    Z((1-\theta)t_1 + \theta t_2 )
    &= W(y((1-\theta)t_1 + \theta t_2)) \\
    &= W((1-\theta)y_1(t) + \theta y_2(t)) \\
    &\geq (1-\theta)W(y_1(t)) + \theta W(y_2(t)) \\
    &= (1-\theta)Z(t_1) + \theta Z(t_2), 
\end{align*}
which means that \( Z(t) \) is concave in \( t \). 

\subsection{Partials}

The partial derivatives of the critical population size \( n_c \) with respect to the dimensionless parameters \( \eta \), \( \nu \), and \( r \) 
are obtained as follows: 
\begin{align*}
    \frac{\partial n_c}{\partial \eta} 
    &= \frac{2r\eta + r + 1}{\nu (r-1)} > 0, \\
    \frac{\partial n_c}{\partial \nu } 
    &= - \frac{(1+\eta)(r\eta + 1)}{\nu^2 (r-1)} < 0, \\
    \frac{\partial n_c }{\partial r} 
    &= - \frac{(1+\eta)^2}{\nu (r-1)^2} < 0.
\end{align*}
We note that \( r > 1 \) by definition.

\section{\label{sec:acknolwedgments} Acknowledgments} 

The authors thank Hiraku Nishimori for discussions on earlier results of this study. H.G.~acknowledges support from the Japan Society for the Promotion of Science (JSPS) under the JSPS Research Fellowship for Young Scholars, from the Japan Science Society under the Sasakawa Scientific Research Grant No.~2024-6036, and from the Meiji University Overseas Challenge Program.

\bibliographystyle{apsrev4-2}
\bibliography{refs}

\end{document}


\title{
Supporting Information for ``When it pays to teach: a population threshold for dedicated teaching'' 
}

\author{Hirotaka Goto}
\affiliation{Graduate School of Advanced Mathematical Sciences, 
Meiji University, 4-21-1 Nakano, Tokyo 164-8525, Japan}

\author{Joshua B. Plotkin}
\affiliation{
Center for Mathematical Biology, University of Pennsylvania, Philadelphia, PA 19104
}
\affiliation{
Department of Biology, University of Pennsylvania, Philadelphia, PA 19104
}

\date{June 30, 2025}

\maketitle


\section{A model of teaching and learning} 

Here we derive \( \taup \) and \( \tauopt^\ast \) given in the main text. 

\subsection{Derivation of \( \taup \)}

The productivity function is given by 
\begin{align*}
    F(\tau; n) = (1-\tau)W(\tau;n), 
    \quad W(\tau;n) = \frac{h_0 + h_1 (\eta + \nu \tau n)}{1 + \eta + \nu \tau n}. 
\end{align*}
We compute the optimal proportion of teachers, \( \tauopt \), by imposing the condition 
\begin{align}
    \frac{\partial F}{\partial \tau} = 0. \label{eq:SI_F_extremum_condition}
\end{align}
Given that 
\begin{align*}
    \frac{\partial F}{\partial \tau} = - W + (1-\tau) \frac{\partial W}{\partial \tau} 
\end{align*}
and 
\begin{align*}
    \frac{\partial W}{\partial \tau} = \frac{(h_1 - h_0)\nu n}{(1 + \eta + \nu \tau n)^2}, 
\end{align*}
solving Eq.~\ref{eq:SI_F_extremum_condition} 
yields the quadratic equation
\begin{align*}
    (\nu n \tau)^2 
    + 2 (\eta + 1) \nu n \tau 
    + \eta^2 + \left(1 + \frac{1}{r}\right) \eta - \left(1 - \frac{1}{r}\right) \nu n + \frac{1}{r} 
    = 0, 
\end{align*}
which has the solution  
\begin{align*}
    \tau 
    = \frac{1}{\nu n} 
    \left[-(1+\eta) 
    \pm \sqrt{\left(1 -\frac{1}{r} \right) \left(1 + \eta + \nu n \right)} \right]. 
\end{align*}
Since \( \tau \) is positive, we obtain 
\begin{align*}
    \taup 
    = \frac{1}{\nu n} 
    \left[-(1+\eta) 
    + \sqrt{\left(1 -\frac{1}{r} \right) \left(1 + \eta + \nu n \right)} \right]. 
\end{align*}

\subsection{Derivation of \( \tauopt^\ast \)}

We solve the equation 
\begin{align}
    \frac{\partial \taup}{\partial n} = 0. \label{eq:SI_taup_extremum_condition}
\end{align}
For convenience, we introduce the following notations: 
\begin{align*}
    \beta = 1 + \eta, \quad \gamma  = 1 - \frac{1}{r},
\end{align*}
which, by definition, satisfy 
\( \beta > 1 \), \( \gamma > 0 \), and \( \beta > \gamma \). 
The solution \( \tau_+ \) can then be expressed as 
\begin{align*}
    \taup 
    = \frac{1}{\nu n} 
    \left[ - \beta + \sqrt{\gamma (\beta + \nu n)} \right]. 
\end{align*}
Solving Eq.~\ref{eq:SI_taup_extremum_condition} leads to 
\begin{align}
    \gamma \left(\beta + \frac{1}{2}\nu n \right)^2 = \beta^2 (\beta + \nu n). \label{eq:SI_n_quadratic_convenient}
\end{align}
Reorganizing Eq.~\ref{eq:SI_n_quadratic_convenient} yields the quadratic equation 
\begin{align*}
    \gamma \nu^2 n^2 - 4 (\beta - \gamma ) \beta \nu n - 4 (\beta - \gamma) \beta^2 = 0,
\end{align*}
whose solution is 
\begin{align*}
    n = \frac{2 \beta}{\gamma \nu} \left[(\beta - \gamma) \pm \sqrt{\beta (\beta - \gamma)} \right]. 
\end{align*}
Provided that \( n \) is positive, 
we find the population size \( n^\ast \) at which \( \tauopt (n) \) becomes maximal: 
\begin{align}
    n^\ast = \frac{2 \beta}{\gamma \nu} \left[(\beta - \gamma) + \sqrt{\beta (\beta - \gamma)} \right]. 
    \label{eq:SI_n_star}
\end{align}
By applying Eqs.~\ref{eq:SI_n_quadratic_convenient} and \ref{eq:SI_n_star} in order, 
we obtain 
\begin{align*}
    \taup (n^\ast) 
    &= \frac{1}{\nu n^\ast} 
    \left[ - \beta + \frac{\gamma}{\beta} \left(\beta + \frac{1}{2}\nu n \right) \right] \\ 
    &= \frac{1}{\nu n^\ast} 
    \left[ - \beta + \gamma + (\beta - \gamma) + \sqrt{\beta (\beta - \gamma)} \right] \\ 
    &= \frac{\sqrt{\beta (\beta - \gamma)}}{\nu n^\ast}. 
\end{align*}
Substituting Eq.~\ref{eq:SI_n_star} into the simplified expression obtained above yields 
\begin{align*}
    \taup (n^\ast) 
    = \frac{\gamma}{2 \beta} \frac{\sqrt{\beta}}{\sqrt{\beta} + \sqrt{\beta - \alpha}},
\end{align*}
which is equivalent to \( \tauopt^\ast \coloneqq \max_{n} \{ \tauopt (n) \} \) given in the main text.

\section{Five extensions}

\begin{figure*}[t]
    \centering
    \includegraphics[width=\linewidth]{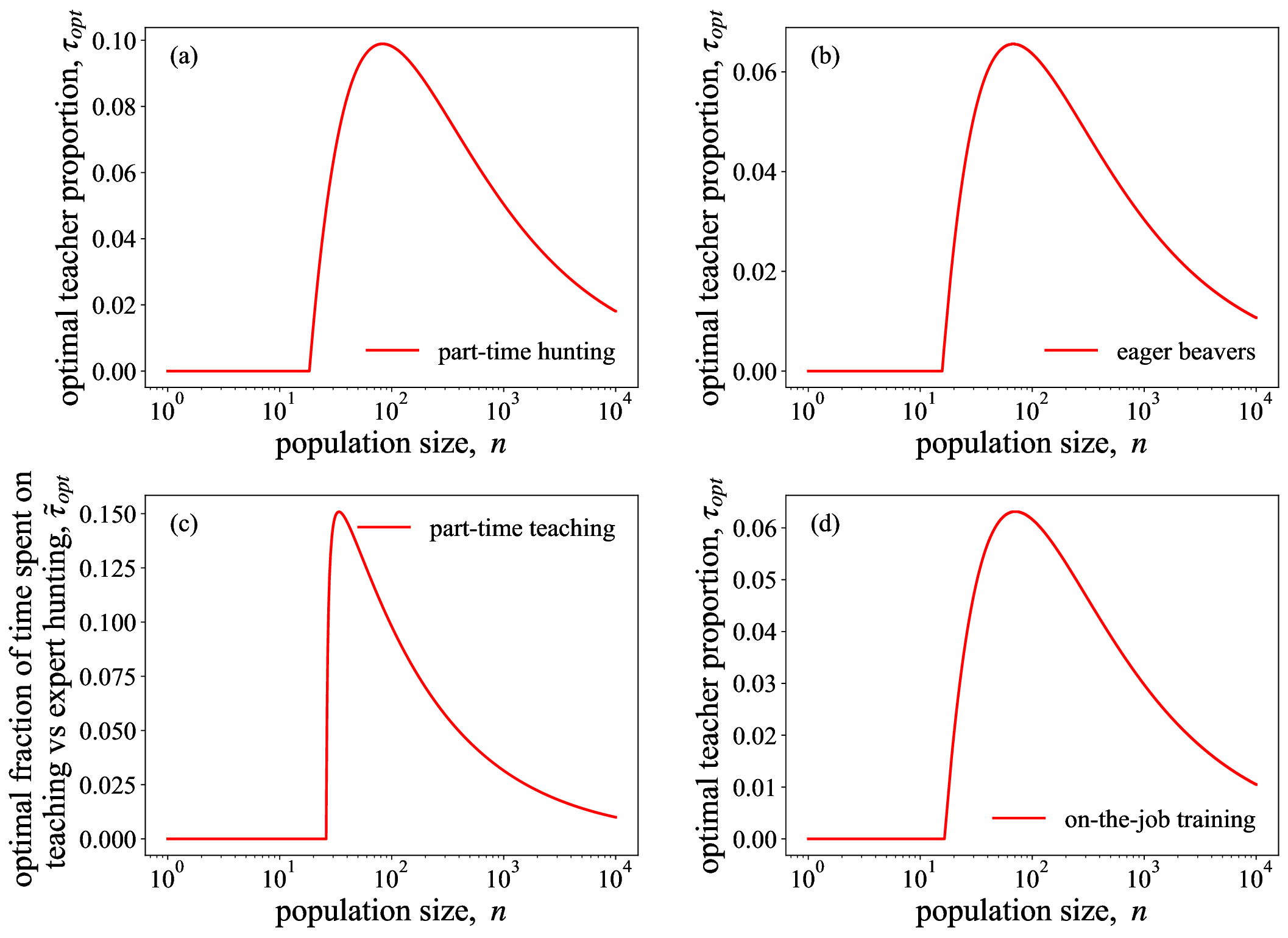}
    \caption{Optimal proportion of teachers or fraction of time dedicated to teaching in our four extended models obtained numerically. 
    Parameters: \( h_0 = 8, h_1 = 10, \lambda_0 = 0.1, \lambda_1 = 1, f = 5 \). 
    (a) part-time hunting (\( \gamma = 0.5, h_T = 9 \)). 
    (b) eager-beavers (\( h_T = 2 \)). 
    (c) part-time teaching. 
    (d) on-the-job training (\( h_T = 0.1 \)). 
    }
    \label{fig:extensions_profiles}
\end{figure*}

We provide full mathematical descriptions of the five extended models mentioned in the main text and present (mostly computational) results regarding the typical behavior of the optimal proportion of teachers for each of these cases (see Fig.~\ref{fig:extensions_profiles} for the first four extensions; see the main text for the finite capacity extension). 
We compute per-capita productivity \( F \) by substituting the equilibrium values of \( n_0 \) and \( n_1 \) obtained from the dynamical equations given in each extension. All hunting rates are positive.

\emph{Part-time hunting}---\emph{teachers can hunt a little bit instead of teaching.} 
Let \( \gamma \) be the proportion of time spent on hunting relative to teaching (\( 0 < \gamma < 1 \)). In other words, teachers engage in hunting (instead of teaching) with a rate of \( 1 - \gamma \). 
We assume that 
\begin{align*}
    n_0 + n_1 = (1-\tau) n, \quad n_T = \tau n. 
\end{align*}
The dynamical equations are given by  
\begin{align*}
    \begin{cases}
        \dot n_0 = f n_1 - \left[\lambda + \mu (1-\gamma) n_T \right] n_0, \\
        \dot n_1 = \left[\lambda + \mu (1-\gamma) n_T \right] - f n_1. \\
    \end{cases}
\end{align*}
The productivity function is then given by 
\begin{align*}
    F = \frac{n_0 h_0 + n_1 h_1 + \gamma n_T h_T}{n}, 
\end{align*}
where we assume \( h_0 < h_T < h_1 \) and \( \gamma \approx 0 \) to be consistent with the fact that hunting is performed only ``part-time.''

\emph{Eager-beavers}---\emph{teachers can hunt a little bit in addition to teaching.} 
In this case, we assume the same basic population structure 
\begin{align*}
    n_0 + n_1 = (1-\tau) n, \quad n_T = \tau n, 
\end{align*}
and the same dynamical equations as in our baseline model, i.e., 
\begin{align*}
    \begin{cases}
        \dot n_0 = f n_1 - (\lambda + \mu n_T ) n_0,  \\
        \dot n_1 =  (\lambda + \mu n_T ) n_0 - f n_1. 
    \end{cases} 
\end{align*} 
However, the productivity function is slightly tweaked to reflect the fact that they both teach and hunt: 
\begin{align*}
    F = \frac{n_0 h_0 + n_1 h_1 + n_T h_T}{n}, 
\end{align*}
where we assume \( h_0 < h_T < h_1 \).

\emph{Part-time teaching}---\emph{expert hunters teach amateur hunters, i.e., there are no dedicated teachers.} 
Here we have a simpler population structure (without a dedicated teaching class): 
\begin{align*}
    n_0 + n_1 = n. 
\end{align*}
We represent the fraction of time expert hunters spend on teaching by \( \tilde\tau \) (\( 0 < \tilde\tau < 1 \)). 
Then, the dynamical equations are 
\begin{align*}
    \begin{cases}
        \dot n_0 = f n_1 - (\lambda + \mu \tilde\tau n_1) n_0, \\
        \dot n_1 = (\lambda + \mu \tilde\tau n_1) n_0 - fn_1,
    \end{cases}
\end{align*}
whereas the productivity function is given by  
\begin{align*}
    F = \frac{n_0 h_0 +  (1-\tilde\tau) n_1 h_1}{n}. 
\end{align*}

\emph{On-the-job training}---\emph{teaching involves a little bit of hunting.} 
Similar to the \emph{eager-beavers} extension, we assume the same population structure and dynamical equations as in our baseline model: 
\begin{align*}
     n_0 + n_1 = (1-\tau) n, \quad n_T = \tau n, 
\end{align*}
and 
\begin{align*}
    \begin{cases}
        \dot n_0 = f n_1 - (\lambda + \mu n_T ) n_0,  \\
        \dot n_1 =  (\lambda + \mu n_T ) n_0 - f n_1. 
    \end{cases} 
\end{align*} 
Because teaching includes a demonstration of hunting, the productivity function involves an additional term: 
\begin{align*}
    F = \frac{n_0 h_0 + n_1 h_1 + \mu n_T n_0 h_T}{n}, 
\end{align*}
where \( h_T < h_0 < h_1 \).

\emph{Finite teaching capacity}---\emph{teachers have a limited teaching capacity.} 
Assuming the same population structure as in our baseline model, 
\begin{align*}
     n_0 + n_1 = (1-\tau) n, \quad n_T = \tau n, 
\end{align*}
our revised model, also presented in the main text, is given by 
\begin{align*}
    \begin{dcases}
        \dot{n}_0 = f n_1 - \left(\lambda + \mu \frac{K}{n + K} n_T \right) n_0, \\
        \dot{n}_1 = \left(\lambda + \mu \frac{K}{n + K} n_T\right) n_0 - f n_1. 
    \end{dcases}
\end{align*}
where \( K \) represents the half-saturation point in teaching capacity, that is, the population size where teaching capacity drops to half of its maximum.  

The numbers of amateur and expert hunters at equilibrium are given by 
\begin{align*}
    (n_0^\ast, \;n_1^\ast) = \frac{(1-\tau)n}{1 +\eta + \nu \tau \chi(n) n} (1, \;\eta + \nu \tau \chi(n) n), 
\end{align*}
where the function \( \chi (n) \) represents the saturation effect: 
\begin{align*}
    \chi(n) = \frac{K}{n + K}.
\end{align*}
Therefore, the ratio of those numbers at equilibrium is computed by 
\begin{align*}
    \frac{n_1^\ast}{n_0^\ast} = \eta + \nu \tau \chi(n)n. 
\end{align*}
In the main text, 
we show that the optimal proportion of teachers, \( \tauopt (n) \), is nonzero unless \( K < n_c \). 
As the population size \( n \) tends to infinity, \( \tauopt \) converges as 
\begin{align*}
    \tauopt 
    \to 
    \frac{ -(1+\eta) + \sqrt{(1 - 1/r) (1+\eta +\nu K)}}{\nu K}. 
\end{align*}
Importantly, this limiting teacher proportion is strictly positive, provided that \( K > n_c \), which marks a crucial difference between the finite-teaching capacity extension and the baseline model. 
Consequently, the ratio of the number of expert hunters \( n_1^\ast \) to that of amateur hunters \( n_0^\ast \), when the number of teachers is optimal, converges to 
\begin{align}
    \lim_{n \to \infty} 
    \left. \frac{n_1^\ast}{n_0^\ast} \right|_{\tau = \tauopt}
    =
    - 1 + \sqrt{\left(1 - \frac{1}{r}\right) (1+\eta +\nu K)}, 
\end{align}
which is also guaranteed to be positive, given that \( K > n_c \). 

\begin{figure}[t]
    \centering
    \includegraphics[width=\columnwidth]{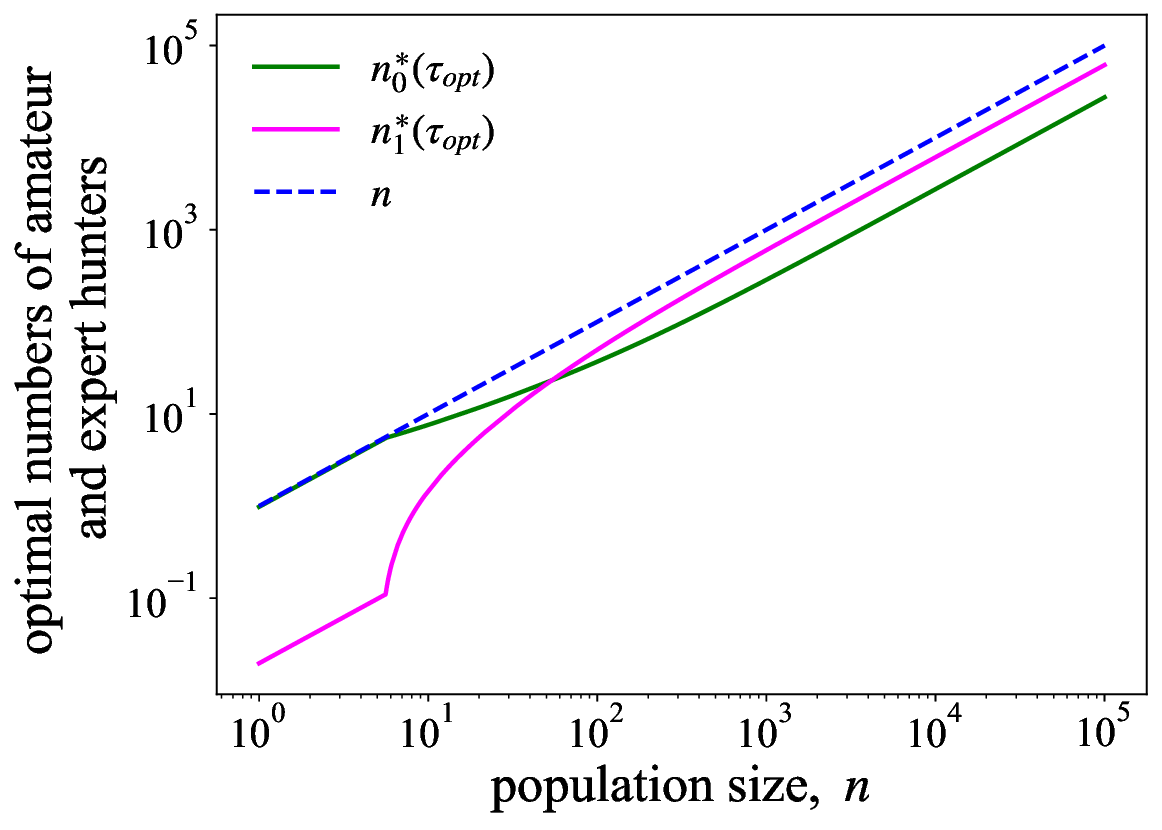}
    \caption{The ratio of expert hunters \( n_1^\ast (\tauopt) \) to amateur hunters \( n_0^\ast (\tauopt) \) does not either diverge or vanish but remains constant in the large-\( n \) limit, which is not the case in the baseline hunter-teacher model (see the main text).}
    \label{fig:finite_capacity_model_optimal_ratio}
\end{figure}

\section{A model with three levels of expertise}

\subsection{Model definition}

Our extended model with three levels of expertise, also briefly presented in the main text, is given by 
\begin{align*}
    \begin{cases}
        \dot{n}_0 = f n_1 - (\lambda_0 + \mu_0 \alpha n_T) n_0, \\
        \dot{n}_1 = (\lambda_0 + \mu_0 \alpha n_T) n_0 - f n_1 + g n_2 \\
        \quad \quad \quad - \left[\lambda_1 + \mu_1 (1-\alpha) n_T\right] n_1, \\
        \dot{n}_2 = \left[\lambda_1 + \mu_1 (1-\alpha) n_T\right] n_1 - g n_2, 
    \end{cases}
\end{align*}
where we define productivity function \( F \) as 
\begin{align*}
    F = \frac{h_0 n_0 + h_1 n_1 + h_2 n_2}{n},
\end{align*}
assuming \( h_0 < h_1 < h_2 \).

\subsection{Overview of the analysis}

We derive the two critical populations sizes, \( n_c \) and \( n_\alpha \), that characterize the emergence of a teaching class within a population and the emergence of differentiated teaching, respectively. 
Our goal is to find \( \tau \) and \( \alpha \) that maximize productivity in a population of size \( n \), which is explicitly given by 
\begin{align*}
    F(\tau, \alpha; n) 
    = (1-\tau) 
    \frac{h_0 w_0 + h_1 w_1 + h_2 w_2}{w_0 + w_1 + w_2}, 
\end{align*}
where the weights \( w_i \) correspond to the hunting rates \( h_i \): 
\begin{align*}
    w_0 &= fg, \\
    w_1 &= g(\lambda_0 + \mu_0 \alpha n \tau), \\
    w_2 &= (\lambda_0 + \mu_0 \alpha n \tau)(\lambda_1 + \mu_1 (1-\alpha) n \tau). 
\end{align*}
The most straightforward way to find such optimal parameter values is computing the following equations simultaneously: 
\begin{align}
    \frac{\partial F}{\partial \tau} = 0, \quad \frac{\partial F}{\partial \alpha} = 0. 
    \label{eq:three_stage model_first_order_conditions}
\end{align}
However, due to algebraic complexity, 
it is highly unlikely to obtain closed-form solutions \( \tauopt \) and \( \alphaopt \). 
In fact, solving \( \partial_\tau F = 0 \) leads to an algebraic equation that is quartic in \( \tau \) and quadratic in \( \alpha \), whereas \( \partial_\alpha F = 0 \) yields another algebraic equation that is quadratic in both parameters. 
Moreover, 
the full expressions of \( \tauopt \) and \( \alphaopt \) may not be necessary if we are interested only in critical population sizes at which the optimal parameter values change abruptly, that is, undergo a phase transition. 
To this end, 
we adopt an ad hoc but more efficient approach. 
From this point onward, we focus only on the regime where a nonzero \( \tauopt \) emerges \emph{continuously} from zero at a critical population size. 
In fact, simulations suggest that \( \tauopt \) \emph{could} exhibit a discontinuous transition in some parameter regime as the population size \( n \) varies, 
which is beyond of the scope of this method.

\subsection{The \emph{first} critical population size \( n_c \)}

Let us introduce 
\begin{align*}
    U(\tau, \alpha; n) &= h_0 w_0 + h_1 w_1 + h_2 w_2, \\
    V(\tau, \alpha; n) &= w_0 + w_1 + w_2, 
\end{align*}
which allows us to rewrite productivity as \( F(\tau, \alpha; n) \equiv (1-\tau) U(\tau, \alpha; n)/V(\tau, \alpha; n) \). 
Reorganizing \( U \) and \( V \) in terms of \( \tau n \) gives 
\begin{align*}
    U(\tau, \alpha; n) 
    &= u_0 + u_1(\alpha) \tau n + u_2(\alpha) (\tau n)^2, \\
    V(\tau, \alpha; n) 
    &= v_0 + v_1(\alpha) \tau n + v_2(\alpha) (\tau n)^2, 
\end{align*}
where 
\begin{align*}
    u_0 &= h_0 fg + h_1 g \lambda_0 + h_2 \lambda_0 \lambda_1, \\
    u_1 &= h_1 g \mu_0 \alpha + h_2 [\lambda_0 \mu_1 (1-\alpha) + \lambda_1 \mu_0 \alpha], \\
    u_2 &= h_2 \mu_0 \mu_1 \alpha (1-\alpha),  \\
    v_0 &= fg + g \lambda_0 + \lambda_0 \lambda_1, \\ 
    v_1 &= g \mu_0 \alpha + \lambda_0 \mu_1 (1-\alpha) + \lambda_1 \mu_0 \alpha, \\
    v_2 &= \mu_0 \mu_1 \alpha (1-\alpha).
\end{align*}
Using the expressions above, 
we expand \( F \) in powers of \( \tau \) as 
\begin{align*}
    F(\tau, \alpha; n) 
    &= F_0 + F_1 \tau + F_2 \tau^2 + O(\tau^3), 
\end{align*}
where 
\begin{align*}
F_0 &= \frac{u_0}{v_0}, \\
F_1 &= \frac{1}{v_0}\left[\left(u_1 - u_0 \frac{v_1}{v_0}\right)n - u_0\right], \\
F_2 &= 
\frac{1}{v_0} \left[\left(-\frac{v_2}{v_0} + \frac{v_1}{v_0^2} - u_1 \frac{v_1}{v_0} + u_2 \right) n^2 - \left(u_1 - u_0 \frac{v_1}{v_0} \right)n \right]. 
\end{align*}
Then, we identify the partial derivatives of \( F \) with respect to \( \tau \): 
\begin{align*}
    \frac{\partial F}{\partial \tau} &= F_1 + O(\tau), \\
    \frac{\partial^2 F}{\partial \tau^2} &= 2 F_2 + O(\tau). 
\end{align*}
We note that for any \( \alpha \in [0,1] \), the partial derivative 
\begin{align*}
    \frac{\partial F}{\partial \tau} (0, \alpha, n) 
    &= F_1(\alpha; n) 
\end{align*}
is positive for large \( n \) but negative for small \( n \). 
By solving 
\begin{align*}
    \frac{\partial F}{\partial \tau}(0, \alpha; n) 
    = 0 
    \quad \Leftrightarrow 
    \quad 
    F_1 = 0, 
\end{align*}
we find 
\begin{align*}
    n_{c} = \frac{u_0 v_0}{u_1 v_0 - u_0 v_1}. 
\end{align*}
Notably, the denominator is positive for any choice of \( \alpha \) because 
\begin{align*}
u_1 v_0 - u_0 v_1 
= (h_1 - h_0) fg^2 \mu_0 \alpha 
+ (h_2 - h_1) g \lambda_0^2 \mu_1 (1-\alpha) \\
+ (h_2 - h_0) fg [\lambda_0 \mu_1(1-\alpha) + \lambda_1 \mu_0 \alpha]. 
\end{align*}
Therefore, 
\( n_c \) is always positive, regardless of the value of \( \alpha \). 
We note, however, that 
its exact value still depends upon \( \alpha \), i.e., \( n_c = n_c (\alpha) \). 
We leave it \emph{undetermined} 
until we come back later and identify its exact value \emph{post hoc}, 
which we will refer to as \( n_c^\ast \).

\subsection{How \( \tauopt \) behaves around \( n_c \)} 

The first-order condition 
\begin{align*}
    \frac{\partial F}{\partial \tau} 
    = F_1 + 2F_2 \tau + O(\tau^2) = 0
\end{align*}
reveals how \( \tauopt \) roughly behaves around the critical point \( n = n_c \), which can be summarized as follows: 
\begin{align*}
    \tau_{opt} 
    &\approx - \frac{F_1}{2 F_2} \propto \Delta n, 
\end{align*}
where we define \( \Delta n = n - n_c > 0 \). 
This implies that nonzero \( \tauopt \) emerges 
\emph{linearly} from \( \tauopt = 0 \).

\subsection{The \emph{second} critical population size \( n_\alpha \)}

We now determine how \( \alphaopt \) behaves around \( n = n_c \). 
Since the transition is continuous and linear, 
it is reasonable to assume that \( \tauopt \) is still small around \( n = n_c \). 
Since \( F_0 \) is independent of \( \alpha \), 
\begin{align*}
    \frac{\partial F}{\partial \alpha} 
    = \frac{\partial F_1}{\partial \alpha} \tau + O(\tau^2).
\end{align*}
Therefore, if \( \tau \) is small, 
the sign of \( \partial_\alpha F \) depends entirely upon 
\( \partial_\alpha F_1 \), 
which can be unpacked as
\begin{align*}
\frac{\partial F_1}{\partial \alpha} 
&= \frac{1}{v_0^2} \left[v_0 \frac{\partial u_1}{\partial \alpha} - u_0 \frac{\partial v_1}{\partial \alpha} \right]n \\
&= \frac{g}{v_0^2}\bigl[(h_1 - h_0) fg + (h_2 - h_0) f (\lambda_1 \mu_0 - \lambda_0 \mu_1) \\
& \quad \quad \quad - (h_2 - h_1) \lambda_0^2 \mu_1 \bigr]n. 
\end{align*}
Assuming that this is positive, 
it immediately follows that 
\( \alphaopt = 1 \) around the critical point. 
To proceed, we impose that \( \alphaopt \neq 1 \) also emerges \emph{continuously} from \( \alphaopt = 1 \). 
Then, we find the exact value of \( n \) where \( \alphaopt \neq 1 \) arises by solving the equation 
\begin{align*}
    \left.\frac{\partial F}{\partial \alpha}\right|_{\alpha=1} = 0, 
\end{align*}
which is equivalent to 
\begin{align*}
    \left.\frac{\partial W}{\partial \alpha}\right|_{\alpha=1} = 0, 
\end{align*}
since \( \tauopt \neq 1 \). 
To begin with, we compute 
\begin{align*}
    \frac{\partial W}{\partial \alpha}(\tau, 1; n)
    &= \left.\frac{\frac{\partial U}{\partial \alpha} V - U \frac{\partial V}{\partial \alpha}}{V^2} \right|_{\alpha=1} \notag \\
    &= 
    \frac{g \tau n }{V(\tau, 1; n)^2} 
    (Q_0 + Q_1 n + Q_2 n^2 )
\end{align*}
where 
\begin{align*}
    Q_0 &= f (\mu_0 \lambda_1 - \mu_1 \lambda_0)( h_2 - h_0 ) \notag \\ 
    &\quad \quad + fg \mu_0 (h_1-h_0) - \mu_1 \lambda_0^2 (h_2-h_1), \\
    Q_1 &= - \mu_0 \mu_1 \left[ 2\lambda_0 ( h_2 - h_1 ) + f ( h_2 - h_0 ) \right] \tau , \\
    Q_2 &= - \mu_0^2 \mu_1 ( h_2 - h_0 ) \tau^2. 
\end{align*}
Note that \( Q_1 < 0 \) and \( Q_2 < 0 \) but \( Q_0 \) can be either positive or negative depending on the parameter values. 
Importantly, it follows that 
\begin{align*}
    Q_0 > 0 \quad \Leftrightarrow \quad \frac{\partial F_1}{\partial \alpha} > 0. 
\end{align*}
Solving \( \partial_{\alpha} W (\tau,1;n)  = 0 \) yields 
\begin{widetext}
\begin{align*}
    n_\alpha(\tau) = \frac{1}{\nu_0 \tau}\left[\frac{-\nu_1 (r_2-1) + \sqrt{\nu_1^2 (r_2-1)^2 + 4\nu_0\nu_1 (r_2-r_1)\left[(r_2-1)\eta_1+(r_1-1)\right]}}{2\nu_1 (r_2-r_1)} - \eta_0 \right], 
\end{align*}
\end{widetext}
under a suitable change of parameters given by 
\begin{align*}
    \eta_0 = \frac{\lambda_0}{f}, \eta_1 = \frac{\lambda_1}{g}, \nu_0 = \frac{\mu_0}{f}, \nu_1 = \frac{\mu_1}{g}, r_1 = \frac{h_1}{h_0}, r_2 = \frac{h_2}{h_0}. 
\end{align*}
The function \( n_\alpha \)
is guaranteed to be positive if and only if \( Q_0 > 0 \). 
Note that \( n_\alpha \propto 1/\tau \) (see the orange curve in Fig.~7 in the main text).

\subsection{Identifying the exact value of \( n_\alpha^\ast \)}

Let us next solve 
\begin{align}
    \frac{\partial F}{\partial \tau}(\tau, 1; n)
    = 0 \label{eq: first order alpha 1}
\end{align}
to find the conditional optimum \( \left.\tauopt \right|_{\alpha=1}(n) \). 
Note that this is different from \( \tauopt (n) \), which is optimized for all \( \alpha \): i.e., \( \left.\tauopt \right|_{\alpha=1}(n) \) is optimal given that \( \alpha = 1 \). 
After some algebraic calculations, 
we find that 
\begin{align*}
    \frac{\partial F}{\partial \tau}(\tau, 1; n)
    = 
    \frac{- a \tau^2 - b \tau - c}{[fg + (g+\lambda_1)(\lambda_0 + \mu_0 \tau n)]^2}, 
\end{align*}
where 
\begin{align}
    a &= M J (\mu_{0}n)^{2}, \nonumber \\
    b &= \left[f g (h_{0}J + M + L) + 2 M J\lambda_{0} \right] \mu_{0} n , \nonumber \\
    c &= h_{0}f^{2}g^{2} + f g (h_{0}J + M)\lambda_{0} + M J \lambda_{0}^{2} - fgL \mu_0 n, \label{eq:SI_c} 
\end{align}
with 
\begin{align*}
    J &= g + \lambda_1, \\
    M &= h_1 g + h_2 \lambda_1, \\
    L &= (h_2 - h_0)\lambda_1 + (h_1 -h_0)g. 
\end{align*} 
Reorganizing the terms in Eq.~\ref{eq:SI_c} reveals that 
\begin{align*}
    c \equiv u_0 v_0 - (u_1 v_0 - u_0 v_1) n, 
\end{align*}
which means that 
\begin{align*}
    c < 0 \quad \Leftrightarrow \quad n > n_c(1), 
\end{align*}
ensuring that Eq.~\ref{eq: first order alpha 1} 
has a real positive solution 
\begin{align*}
    \left.\taup \right|_{\alpha=1} = \frac{-b + \sqrt{b^2-4ac}}{2a}, 
\end{align*}
provided \( n > n_c(1)\); 
otherwise, \( \left.\tauopt \right|_{\alpha=1} = 0 \) because \( \partial_\tau F (\tau, 1; n) < 0 \) for all \( \tau \in [0,1] \). 
To sum up, we have obtained 
\begin{align*}
    \left.\tauopt \right|_{\alpha=1}
    = 
    \begin{cases}
        0 \quad &n < n_c(1), \\
        \left.\taup \right|_{\alpha=1} \quad &n > n_c(1), 
    \end{cases}
\end{align*}
which corresponds to the purple curve in Fig.~7 in the main text.

\subsection{Identifying the exact value of \( n_c^\ast \)}

Since we have derived the explicit functional forms of both 
\( \left.\tauopt \right|_{\alpha=1}(n) \) and \( n_{\alpha}(\tau) \), 
their intersection point---which must be greater than \( n_c(1) \) by the clear geometrical constraint---finally pins down the \emph{second} critical population size \( n_\alpha^\ast \) 
(note that \( n_c(\alpha) < n_\alpha^\ast \) actually holds for any choice of \( \alpha \)). 
Since \( \alpha = 1 \) is optimal below the curve \( n_\alpha (\tau)\) 
whereas \( \alpha < 1 \) is optimal above the curve---provided that there are a positive number of teachers, i.e., \( \tau > 0 \)---it follows that \( n_c^\ast \) must equal \( n_c(1) \).

\subsection{Summary of the results regarding \( n_c^\ast \) and \( n_\alpha^\ast \)}

The results obtained above are summarized as follows: 
\begin{align*}
    \tauopt 
    &= 
    \begin{cases}
        0 \quad & n \leq n_c^\ast, \\
        \left. \taup \right|_{\alpha=1} \quad & n_c^\ast \leq n \leq n_\alpha^\ast, \\
        \tauopt^\ast \quad & n_\alpha^\ast \leq n , 
    \end{cases} 
    \\
    \alphaopt 
    &= 
    \begin{cases}
        0 \quad\quad & n \leq n_c^\ast, \\
        1 \quad\quad & n_c^\ast \leq n \leq n_\alpha^\ast, \\
        \alphaopt^\ast \quad\quad & n_\alpha^\ast \leq n , 
    \end{cases}
\end{align*}
where both \( \tauopt^\ast \) and \( \alphaopt^\ast \) represent optimal values as a function of population size \( n \) that arise from solving Eq.~\ref{eq:three_stage model_first_order_conditions} directly (note however that those functions cannot be explicitly identified using our method presented above).  
The main text includes an illustrative figure (Fig.~7) where the orange and purple lines represent \( n_{\alpha}(n) \) and \( \left.\tauopt \right|_{\alpha=1}(n) \), respectively. 
Their intersection point correctly identifies the critical population size \( n_{\alpha}^\ast \), at which the red curve \( \tauopt \) changes its slope abruptly. 
Additionally, \( n_c^\ast = n_c(1) \) also correctly predicts the first critical population size.